\documentclass[trackchanges,twocolumn]{aastex701}

\usepackage{CJKutf8}
\usepackage{amsmath}

\usepackage{amsmath}

\begin{document}

\newcommand{\lco}{$L'_{\mathrm{CO}}$}
\newcommand{\fH}{f$_{H_2}$}

\newcommand{\sig}{$\sigma_{\rm \ast}$}
\newcommand{\siggas}{$\sigma_{\rm gas}$}

\newcommand{\co}{$^{12}$CO(1-0)}
\newcommand{\coii}{$^{12}$CO(2-1)}

\newcommand{\nuvr}{NUV-$r$}
\newcommand{\mh}{$M_{H_2}$}
\newcommand{\mhi}{$M_{HI}$}
\newcommand{\mstar}{$M_{\rm \ast}$}
\newcommand{\must}{$\mu_{\ast}$}
\newcommand{\hmol}{$H_2$}
\newcommand{\rmol}{$R_{mol}$}
\newcommand{\tdep}{$t_{dep}({\rm H_2})$}
\newcommand{\tdepHI}{$t_{dep}({\rm HI})$}
\newcommand{\fgas}{$f_{\rm H_2}$}
\newcommand{\fhi}{$f_{\rm HI}$}
\newcommand{\xco}{$\alpha_{CO}$}
\newcommand{\ms}{SFR-$M_{\ast}$}
\newcommand{\deltams}{$\Delta$(MS)}
\newcommand{\ntot}{532}
\newcommand{\hi}{{H{\sc i}}}


\newcommand{\xh}{$<\chi_{H I}>$}
\newcommand{\TQ}{$t_{Q}$}

\newcommand{\paa}{{Pa$\alpha$}}
\newcommand{\pab}{{Pa$\beta$}}

\newcommand{\av}{{A$_{V}$}}
\newcommand{\ebv}{E(B$-$V)}

\newcommand{\siv}{[S~{\sc iv}] 10.51 $\mu$m}
\newcommand{\oiiitext}{[O~{\sc iii}]}
\newcommand{\oiiihb}{[O~{\sc iii}]/H$\beta$}

\newcommand{\sivtext}{[S~{\sc iv}]}

\newcommand{\lya}{Ly$\alpha$}

\newcommand{\Lbol}{$L_{\rm bol}$}
\newcommand{\mbh}{$M_{\rm BH}$}
\newcommand{\edd}{$\lambda_{\rm Edd}$}
\newcommand{\logLbol}{log($L_{\rm bol}$/$L_{\sun}$)}

\newcommand{\fwhmhb}{FWHM$_{\rm H\beta}$}
\newcommand{\fwhmha}{FWHM$_{\rm H\alpha}$}

\newcommand{\cii}{[C~{\sc ii}] 158 \mum}
\newcommand{\ciitext}{[C~{\sc ii}]}

\newcommand{\mum}{\ifmmode{\rm \mu m}\else{$\mu$m}\fi}
\newcommand{\vdisp}{$\vdisp$}
\newcommand{\wba}{W$_{80}$}
\newcommand{\wjiu}{W$_{90}$}
\newcommand{\vwu}{{$v_{50}$}}
\newcommand{\vjiu}{{v$_{90}$}}
\newcommand{\vyi}{{v$_{10}$}}
\newcommand{\vbasi}{{v$_{84}$}}
\newcommand{\vyiliu}{{v$_{16}$}}
\newcommand{\flux}{erg cm$^{-2}$ s$^{-1}$}
\newcommand{\fsb}{erg cm$^{-2}$ s$^{-1}$ }

\newcommand{\Lwu}{{$\lambda L_{\lambda}(5100)$}}

\newcommand{\msigma}{$M_{\rm BH}$--$\sigma_\ast$}

\newcommand{\vjiuba}{{$v_{98}$}}
\newcommand{\vlingwu}{{$v_{05}$}}
\newcommand{\vjiuwu}{{$v_{95}$}}
\newcommand{\ajiuyi}{{$A_{91}$}}
\newcommand{\ajiuyiha}{{$A_{91,H\alpha}$}}
\newcommand{\ajiuyio}{{$A_{91,[O~\sc{III}]}$}}

\newcommand{\ewo}{ EW$_{\rm [O{\scriptsize III}]}$ }
\newcommand{\ewfe}{ EW$_{\rm Fe{\scriptsize II}}$ }
\newcommand{\ewhb}{ EW$_{\rm H{\beta}, broad}$ }
\newcommand{\RFe}{ R$_{\rm Fe}$ }

\newcommand{\spi}{{\it Spitzer}}
\newcommand{\her}{{\it Herschel}}
\newcommand{\oi}{\hbox{[O$\,${\scriptsize I}]}}

\newcommand{\oiitext}{{[O$\,${\scriptsize II}]}}
\newcommand{\oii}{{[O$\,${\scriptsize II}] $\lambda$$\lambda$3726,3729}}
\newcommand{\oiii}{{[O$\,${\scriptsize III}] $\lambda$5007}}
\newcommand{\oiiiab}{{[O$\,${\scriptsize III}] $\lambda$$\lambda$4959,5007}}
\newcommand{\oiiib}{{[O$\,${\scriptsize III}] $\lambda$4959}}

\newcommand{\oiiicc}{{[O$\,${\scriptsize III}] $\lambda$4363}}

\newcommand{\nv}{\hbox{N$\,${\scriptsize V}}}
\newcommand{\civ}{\hbox{C$\,${\scriptsize IV} $\lambda$1549}}
\newcommand{\civtext}{\hbox{C$\,${\scriptsize IV}}}

\newcommand{\ciiitext}{\hbox{C$\,${\scriptsize III}]}}
\newcommand{\siivtext}{\hbox{Si$\,${\scriptsize IV}]}}

\newcommand{\nev}{\hbox{[Ne$\,${\scriptsize V}]}}
\newcommand{\niitext}{\hbox{[N$\,${\scriptsize II}]}}
\newcommand{\siitext}{\hbox{[S$\,${\scriptsize II}]}}
\newcommand{\nii}{\hbox{[N$\,${\scriptsize II}] $\lambda$$\lambda$6548,6583}}
\newcommand{\sii}{\hbox{[S$\,${\scriptsize II}] $\lambda$$\lambda$6716,6731}}

\newcommand{\ha}{\hbox{H$\alpha$}}
\newcommand{\hb}{\hbox{H$\beta$}}
\newcommand{\hg}{\hbox{H$\gamma$}}
\newcommand{\hd}{\hbox{H$\delta$}}
\newcommand{\mgii}{\hbox{Mg$\,${\scriptsize II}} $\lambda$2800}
\newcommand{\mgiitext}{\hbox{Mg$\,${\scriptsize II}}}
\newcommand{\feii}{\hbox{Fe$\,${\scriptsize II}}}
\newcommand{\red}[1]{\textcolor{red}{#1}}
\newcommand{\blue}[1]{\textcolor{blue}{#1}}
\newcommand{\kms}{km s$^{-1}$} 
\newcommand{\msun}{$M_{\odot}$} 
\newcommand{\msunyr}{{M$_{\sun}$ yr$^{-1}$}}
\newcommand{\lsun}{\ensuremath{\mathrm{L}_{\odot}}}

\newcommand{\eden}{cm$^{-3}$} 
\newcommand{\momfluxsfr}{$\dot{P}_{SFR}$ }
\newcommand{\momfluxagn}{$\dot{P}_{QSO}$ }
\newcommand{\momfluxout}{$\dot{P}_{outflow}$ }
\newcommand{\momfluxratio}{$\frac{\dot{P}_{outflow}}{\dot{P}_{AGN}}$}
\newcommand{\ergs}{erg\,s$^{-1}$}
\newcommand{\ergscm}{erg\,s$^{-1}$\,cm$^{-2}$}
\newcommand{\myr}{M$_\odot$~yr$^{-1}$} 
\newcommand{\loghn}{log(\nii/\ha) }
\newcommand{\logohb}{log(\oiii/\hb) }

\newcommand\ifsfit{\texttt{IFSFIT}}
\newcommand\lmfit{\texttt{lmfit}}
\newcommand\mpfit{\texttt{mpfit}}
\newcommand\qtdfit{\texttt{q3dfit}}
\newcommand\questfit{\texttt{questfit}}
\newcommand\ppxf{\texttt{ppxf}}

\title{A $z \sim$ 6.2 Quasar on the Local M$_{\rm BH}$--$\sigma_{\rm \ast}$ Relation Quenching Its Host Galaxy from the \textit{Aether} Survey}

\author[0000-0003-3762-7344]{Weizhe Liu \begin{CJK}{UTF8}{gbsn}(刘伟哲)\end{CJK}}
\affiliation{Steward Observatory, University of Arizona, 933 N. Cherry Ave., Tucson, AZ 85721, USA}
\email{wzliu@arizona.edu}

\correspondingauthor{Weizhe Liu}
\email{wzliu@arizona.edu}


\author[0000-0002-6822-2254]{Emanuele Paolo Farina}
\affiliation{International Gemini Observatory/NSF NOIRLab, 670 N A’ohoku Place, Hilo, Hawai'i 96720, USA}
\email{}

\author[0000-0003-3310-0131]{Xiaohui Fan}
\affiliation{Steward Observatory, University of Arizona, 933 N. Cherry Ave., Tucson, AZ 85721, USA}
\email{}

\author[0000-0002-2662-8803]{Roberto Decarli}
\affil{INAF -- Osservatorio di Astrofisica e Scienza dello Spazio di Bologna, via Gobetti 93/3, I-40129, Bologna, Italy }
\email{}

\author[0000-0003-2984-6803]{Masafusa Onoue}
\affiliation{Waseda Institute for Advanced Study (WIAS), Waseda University, 1-21-1, Nishi-Waseda, Shinjuku, Tokyo 169-0051, Japan}
\affiliation{Kavli Institute for the Physics and Mathematics of the Universe (Kavli IPMU, WPI), UTIAS, Tokyo Institutes for Advanced Study, University of Tokyo, Chiba, 277-8583, Japan}
\email{masafusa.onoue@aoni.waseda.jp}

\author[0000-0002-4321-3538]{Haowen Zhang}
\affiliation{Canadian Institute for Theoretical Astrophysics, University of Toronto, Toronto, ON M5S 3H8}
\email{}

\author[0000-0002-2931-7824]{Eduardo Ba\~nados}
\affil{Max Planck Institut f\"ur Astronomie, K\"onigstuhl 17, D-69117, Heidelberg, Germany}
\email{}

\author[0000-0002-3026-0562]{Aaron J. Barth}
\affil{Department of Physics and Astronomy, University of California, Irvine, CA 92697, USA}
\email{}

\author[0000-0002-4770-6137]{Fabrizio Arrigoni Battaia}
\affiliation{
Max Planck Institut f\"ur Astrophysik, Karl Schwarzschild Stra\ss e 1, D-85741 Garching, Germany}
\email{}

\author[0000-0002-4314-021X]{Manuela Bischetti}
\affiliation{Dipartimento di Fisica Enrico Fermi, Università di Pisa, Largo Bruno Pontecorvo 3, Pisa, I-56127, Italy}
\affiliation{INAF-Osservatorio Astronomico di Trieste, Via G. B. Tiepolo 11, I-34131 Trieste, Italy}
\email{manuela.bischetti@unipi.it}

\author[0000-0001-8582-7012]{Sarah E. I. Bosman}
\affiliation{Institute for Theoretical Physics Heidelberg University,
Philosophenweg 12, Heidelberg, D-69117,Germany.}
\affil{Max Planck Institut f\"ur Astronomie, K\"onigstuhl 17, D-69117, Heidelberg, Germany}
\email{}
\email{}

\author[0000-0002-3173-1098]{Hyunseop Choi}
\affiliation{Department of Astronomy, University of Michigan, 1085 S. University Ave., Ann Arbor, MI 48109, USA}
\email{}

\author[0000-0002-6748-2900]{Tiago Costa}
\affiliation{School of Mathematics, Statistics and Physics, Newcastle University, Newcastle upon Tyne, NE1 7RU, UK}
\email{}

\author[]{Simona Gallerani} 
\affiliation{Scuola Normale Superiore, Piazza dei Cavalieri 7, I-56126 Pisa, Italy}
\email{}

\author[0009-0009-8274-441X]{Anniek J. Gloudemans}
\affiliation{NSF NOIRLab, Gemini Observatory, 670 N A'ohoku Place Hilo, HI 96720, USA}
\email{}

\author[0000-0002-5768-738X]{Xiangyu Jin}
\affiliation{Department of Astronomy, University of Michigan, 1085 S. University Ave., Ann Arbor, MI 48109, USA}
\email{}

\author[0000-0002-8858-6784]{Federica Loiacono}
\affiliation{INAF -- Osservatorio di Astrofisica e Scienza dello Spazio di Bologna, via Gobetti 93/3, I-40129, Bologna, Italy}
\email{federica.loiacono1@inaf.it}

\author[0000-0001-5063-0340]{Yoshiki Matsuoka} 
\affiliation{Research Center for Space and Cosmic Evolution, Ehime University, Matsuyama, Ehime 790-8577, Japan}
\email{}

\author[0000-0002-5941-5214]{Chiara Mazzucchelli}
\affiliation{Instituto de Estudios Astrofísicos, Facultad de Ingeniería y Ciencias, Universidad Diego Portales, Avenida Ejercito Libertador 441, Santiago, Chile}
\email{}

\author[0000-0003-4793-7880]{Fabian Walter}
\affil{Max Planck Institut f\"ur Astronomie, K\"onigstuhl 17, D-69117, Heidelberg, Germany}
\affiliation{California Institute of Technology, Pasadena, CA 91125, USA}
\email{}

\author[0000-0002-7633-431X]{Feige Wang}
\affiliation{Department of Astronomy, University of Michigan, 1085 S. University Ave., Ann Arbor, MI 48109, USA}
\email{}

\author[0000-0001-5287-4242]{Jinyi Yang}
\affiliation{Department of Astronomy, University of Michigan, 1085 S. University Ave., Ann Arbor, MI 48109, USA}
\email{}

\begin{abstract}
We report JWST/NIRSpec integral field unit (IFU) observations of the
quasar J1512$+$4422 at $z \sim 6.2$ from the \textit{Aether} survey. 
At $\sim$900 Myr after the Big Bang, this object already lies on the $M_{\rm BH}$--$\sigma_\ast$ relation found in the local universe, with an $M_{\rm BH} \simeq 8.9\times10^8\,M_\odot$ and a stellar velocity dispersion $\sigma_\ast \simeq 288$ km s$^{-1}$. 
We detect an outflow with a velocity of $\sim$478 km s$^{-1}$ in the nuclear region, which likely extends to $\sim$3.2 kpc in projection and has a median velocity of $\sim$352 km s$^{-1}$. The outflow dynamical time scale ($\sim$ 9 Myr) is consistent with the time scale of the current quenching process based on the star formation history as reported previously. The total mass outflow rate (92.6$^{+92.6}_{-74.1}$ M$_{\sun}$ yr$^{-1}$) is larger than the current star formation rate (0.9$^{+3.8}_{-0.8}$ or 4.3$^{+5.8}_{-3.7}$ M$_{\sun}$ yr$^{-1}$), and the total kinetic energy outflow rate (0.6$^{+0.6}_{-0.5}$\% of quasar luminosity) meets the threshold for negative quasar feedback as suggested by simulations. These results suggest that the outflow is capable of suppressing/quenching the star formation activity within the host galaxy. Furthermore, J1512$+$4422 exhibits $\sigma_\ast$, stellar mass and size similar to those of $z \gtrsim$ 3 quiescent/post-starburst galaxies, implying a link between the two. 
Overall, for objects like J1512$+$4422, the evolution of their SMBHs and host galaxies appears to be tightly coupled within the first billion years. The quasar feedback likely plays a critical role in both placing them on the $M_{\rm BH}$--$\sigma_\ast$ relation and quenching. 
\end{abstract}

\section{Introduction}

In the local universe, tight correlations between the masses of supermassive black holes (SMBHs) and their host galaxy properties, such as the stellar velocity dispersion ($\sigma_{\rm \ast}$) and stellar mass ($M_{\rm \ast}$), suggest that the evolution of SMBHs and galaxies are coupled with each other \citep{Ferrarese2000,Gebhardt2000,KormendyHo2013}. Nevertheless, when and how these correlations are established and how SMBHs and galaxies interact with each other across cosmic time remain open questions.

Quasars with $\sim10^{9}-10^{10}\,M_\odot$ SMBHs already exist at $z>6$ \citep[e.g.,][]{Wu2015,Banados2018,Matsuoka2019b,Onoue2019,Shen2019,Yang2020b,Schindler2020,Wang2021,Farina2022,Yang2021,Fan2023,Mazzucchelli2023}, within a billion years after the Big Bang. 
The locations of these quasars on the $M_{\rm BH}$--$\sigma_\ast$ plane offer a powerful probe of the feeding and feedback processes of SMBHs in the early universe and provide tight constraints on models of SMBH and galaxy formation and evolution.
Nevertheless, measuring $\sigma_\ast$ in $z>6$ quasar host galaxies is extremely challenging due to the faintness of the host galaxies and overwhelming glare from the quasar. Gas kinematics based on emission lines like [C\,{\sc ii}] 158 \mum\ has been adopted to trace the host galaxy dynamic mass ($M_{\rm dyn}$) and to probe the $M_{\rm BH}$--$M_{\rm dyn}$ relation. Such studies generally suggest overmassive SMBHs with respect to the local relations \citep[e.g.,][]{Neeleman_kinematics_2021, Willott2017,Izumi2018}, but at least one object already lying on the local relation has been reported \citep[e.g.,][]{Izumi2021}. Alternatively, stellar masses have been measured for $z>$ 6 quasar host galaxies with JWST data \citep[e.g.,][]{Ding2023,Yue2023,Stone2023}.
Again, many of these studies indicate that these high-$z$ quasars lie above the local relations with overmassive SMBHs, while others suggest that some of them may already lie on the local $M_{\rm BH}$--$M_\ast$ relation \citep{Ding2023,Ding2025ApJ,Silverman2025,Ziparo2026}. \citet{Onoue2025}, for the first time, report \sig\ of two $z>6$ quasar host galaxies adopting JWST/NIRSpec fixed-slit (FS) spectroscopy with moderate spectral resolution (R$\sim$1000). Together with their JWST/NIRCam imaging data, they find that one object lies on the local $M_{\rm BH}$--$M_\ast$ relation, while the other hosts an overmassive SMBH \citep[see also][for more discussions on the full SHELLQs JWST Cycle 1 sample where the two objects belong]{Ding2025ApJ,Silverman2025}.

In addition, previous observations have identified a large number of quiescent/post-starburst galaxies at $z \gtrsim 3$ \citep[e.g.,][]{Labbe2005, Glazebrook2017,
Valentino2020,Valentino2023,Alberts2023,Carnall2023,Carnall2024,Ji2024,Nanayakkara2024,Graaff2025,Ito2025}. Their rapid quenching challenges our current understanding of galaxy evolution. Quasar/Active Galactic Nucleus (AGN) feedback is one of the major mechanisms responsible for galaxy quenching. Some simulations indeed suggest that such feedback plays a critical role in shutting down star formation in early galaxies \citep{Lovell2023,Hartley2023}. As a major form of feedback, quasar-driven ionized \citep[e.g.,][]{Bischetti2022,Marshall2023,Marshall2025,Yang2023b,yue_eiger_2023,Loiacono2024,Decarli2024,Liu2024b,Liu2026a,Liu2026b} and molecular \citep[e.g.,][]{Bischetti2019,Salak2024,Spilker2025,zhu2026} outflows have indeed been observed in the early universe. More intriguingly, a recent study discovers that the frequency of fast, galaxy-scale outflows is significantly higher in $z\sim$ 5--6 luminous quasars than the lower-redshift ones, suggesting that quasar feedback may easily suppress/regulate star formation activities in early galaxies \citep{Liu2026a}. Consistently, the two $z\sim6$ quasar host galaxies examined in \citet{Onoue2025}, HSC J151248.71$+$442217.5 and HSC J223644.58$+$003256.9 (hereafter J1512$+$4422 and J2236$+$0032), are rapidly quenching, providing excellent laboratories for a close investigation of the connection between quasar feedback and quenching in the early universe.

In this paper, we report results from the JWST/NIRSpec integral field unit (IFU) observation of one of the two objects from \citet{Onoue2025}, J1512$+$4422 at $z \simeq$ 6.18, as part of the JWST \textit{Aether} survey (ID 5645; PI E. Farina).  J1512$+$4422 was discovered in \citet{Matsuoka2019}. \citet{Onoue2025} suggests that it is quenching rapidly with a specific star formation rate (sSFR) of $\lesssim$0.2 Gyr$^{-1}$ over the last 10 Myr \citep[see also][for more JWST/NIRCam imaging and NIRSpec/FS spectroscopy studies on this object]{Ding2025ApJ,Phillips2025}. We provide new measurements of its \sig\ and other quasar/host galaxy properties and a first spatially-resolved view of the outflow and feedback within this object based on our IFU data. 
We describe the observations and data reduction in Section \ref{2} and the data analysis in Section \ref{3}. We discuss our key findings in Sections \ref{4}, \ref{5} and \ref{6} and summarize them in Section \ref{7}. Throughout the paper, we assume a $\Lambda$CDM cosmology with $H_0 =$ 70 km s$^{-1}$ Mpc$^{-1}$, $\Omega_{\rm m}$ = 0.3, and $\Omega_{\rm \Lambda} = 0.7$. This gives $\sim$5.6 kpc per arcsecond at the redshift of J1512$+$4422.

\section{Observation, Data Reduction}\label{2}

J1512$+$4422 was observed on 2025 April 2 through a JWST survey program named \textit{Aether} survey with the NIRSpec/IFU \citep{Bok2022, Jak2022}. A full description of the entire \textit{Aether} survey will be presented in Farina et al.(in prep.) and Decarli et al. (in prep.). The IFU observation has a field of view (FOV) of $\sim 3\arcsec \times 3 \arcsec$. We adopted a grating/filter configuration of G395H/F290LP, a NRSIRS2 readout pattern with 9 Groups and 1 Integration and 4 SMALL CYCLING dither positions for a total exposure time of $\sim$45 minutes on target. The wavelength coverage is $\sim$2.87--5.14~$\mu$m and the grating has a nominal resolving power $\lambda / \Delta\lambda \simeq 2700$ (velocity resolution FWHM $\sim 130$ \kms). 
The IFU data were reduced following the same procedure as described in \citet{Liu2024b,Liu2026a}. In brief, the data were processed primarily with the JWST Science Calibration Pipeline (version ``1.14.0'' and context file ``jwst\_1293.pmap''). Customized scripts were adopted to replace or improve certain steps in the pipeline, which includes the 1/$f$ noise subtraction in count rate images, the production of the final data cube with the flux-conserving “reproject\_exact” routine in \verb|reproject| and ``EMSM-weighting''\footnote{See Section ``Weighting'' in \url{https://jwst-pipeline.readthedocs.io/en/latest/jwst/cube_build/main.html}}, and the removal of outliers in the final data cube with sigma-clipping. The final combined data cube has a spatial pixel scale of 0\farcs05.

In addition, J1512$+$4422 was also observed with JWST/NIRCam with F150W and F356W filters and JWST/NIRSpec FS spectrograph adopting G395M grating with a nominal resolving power of $\lambda / \Delta\lambda \simeq 1000$ (velocity resolution FWHM $\sim 300$ \kms) and an on-source integration time of $\sim$2.19 hours through JWST program GO \#1967 (PI: M. Onoue). More details of these observations can be found in \citet{Onoue2025}.

\section{Analysis}\label{3}

\subsection{Nuclear Spectrum}
\label{sec:31}

\begin{figure*}[!htb]
    \centering
    \begin{minipage}{1\textwidth}
    \includegraphics[width=\linewidth]{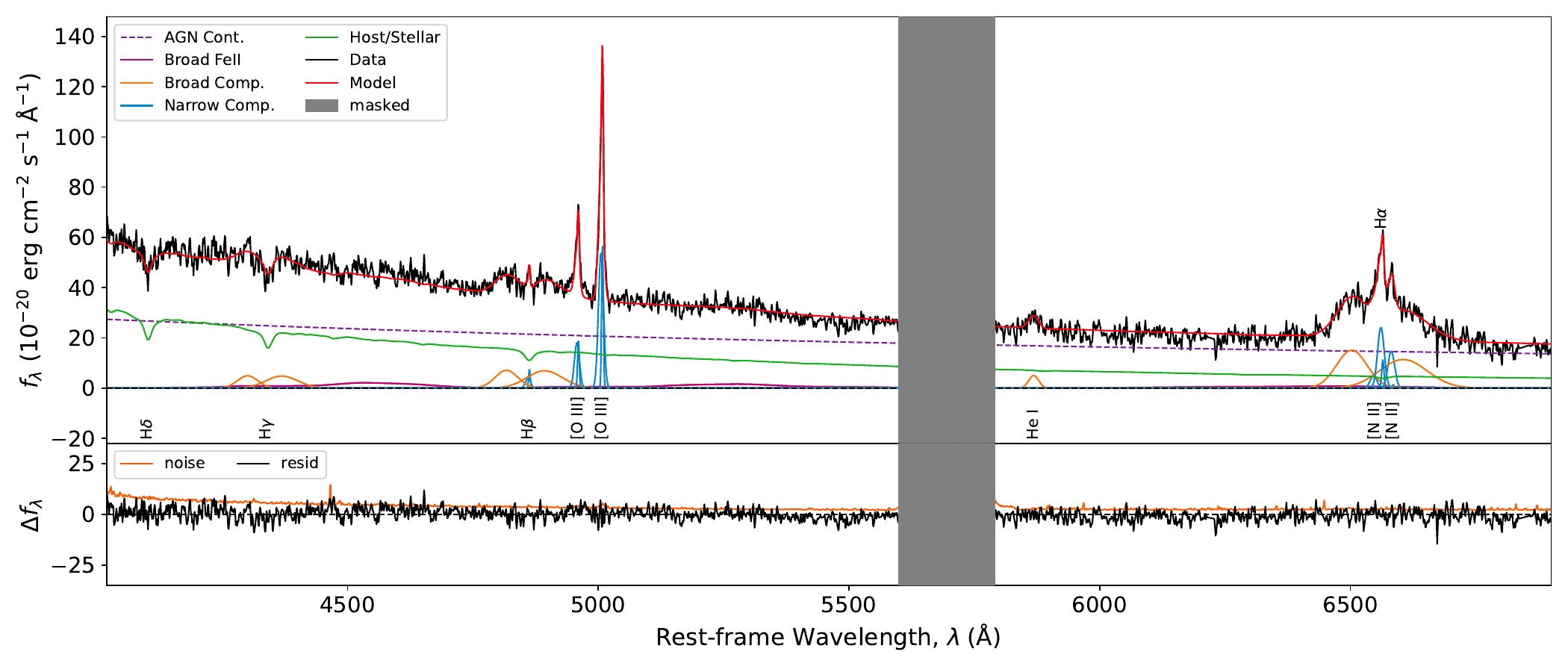}
    \end{minipage}
    \begin{minipage}{0.32\textwidth}
        \centering
    \includegraphics[width=\linewidth]{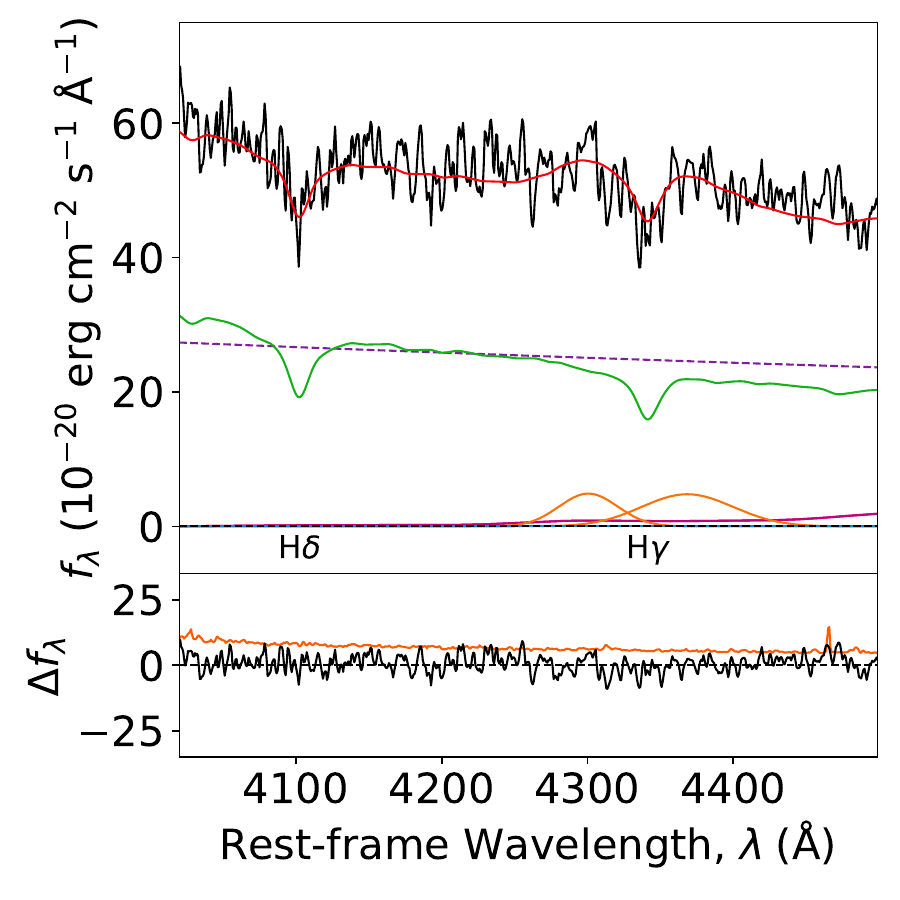}
    \end{minipage}
    \begin{minipage}{0.32\textwidth}
        \centering
    \includegraphics[width=\linewidth]{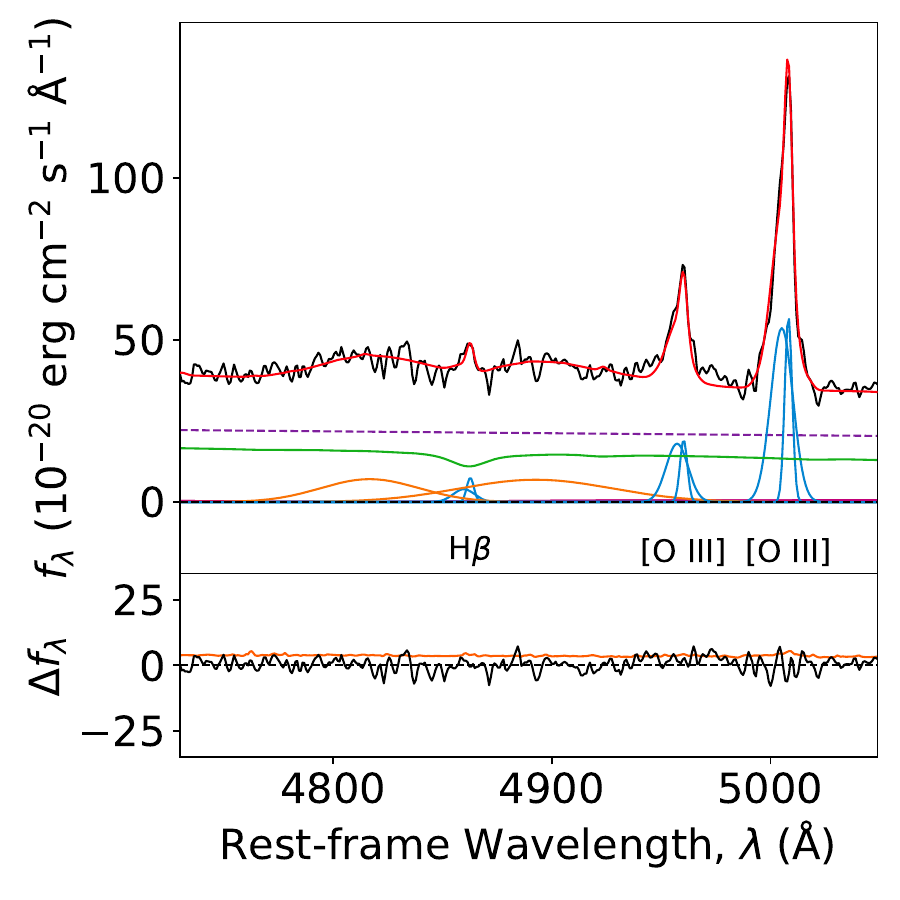}
    \end{minipage}
    \begin{minipage}{0.32\textwidth}
        \centering
    \includegraphics[width=\linewidth]{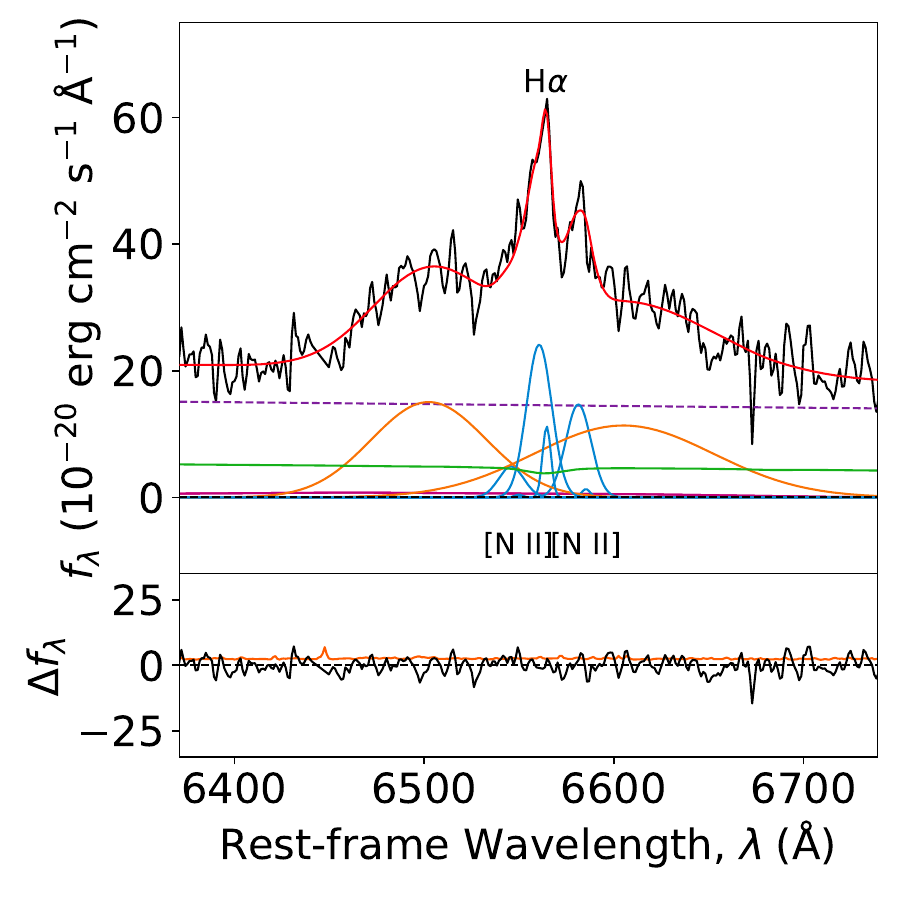}
    \end{minipage}
    \caption{\textbf{Upper Panel:} \textbf{Top:} The nuclear spectrum (black) and best-fit model of J1512$+$4422. The overall best-fit model is shown in red. Also shown are the best-fit power-law continuum (AGN Cont.), \ion{Fe}{2} emission, narrow emission lines (including the systemic and blueshifted components), broad emission lines from the quasar, and host galaxy stellar continuum and absorption lines, as indicated by the legends. The locations of major line features are denoted with their names. The detector gap is masked by the gray shade. \textbf{Bottom:} Residual (black) and spectral uncertainties (orange). \textbf{Lower Panels:} Spectra zoomed into key spectral features.}
    \label{fig:nuclear}
    
\end{figure*}

We fit the nuclear spectrum extracted from a circular aperture with $r=$ 0\farcs1 from the IFU data cube with the public software BADASS \citep{BADASS}. The aperture size is chosen to maximize the S/N of the absorption lines. BADASS is an open-source spectral analysis tool designed for detailed decompositions of Type 1 AGN. It utilizes the Bayesian affine-invariant Markov-Chain Monte Carlo (MCMC) sampler for robust parameter and uncertainty estimation. Readers are referred to their paper for more details on the design of the code.

Our fitting model consisted of a quasar power-law continuum, quasar \feii\ templates from \citet[][]{VC04}, stellar continuum and absorption lines, and broad and narrow emission lines. The scaling and centroid velocity of the \feii\ templates are set as free parameters, and the velocity dispersions of them are allowed to vary freely between [500, 5000] \kms. We have also tested other popular \feii\ templates \citep[e.g.,][]{Boroson1992,Park2022} but found negligible difference in our final best-fits. In the fitting, the models are further convolved with the line spread function of the NIRSpec/IFU G395H grating.

Due to the degeneracy between the quasar continuum and stellar continuum in the fitting, we need to first obtain their relative strength independently. Therefore, we used GALFIT \citep{galfit} to decompose the pure continuum images obtained from the IFU data (see Appendix \ref{galfit} for details) and derived the slope and scale of the quasar power-law continuum which were then fixed in the spectral fitting. 
The stellar continuum and absorption lines were modeled using the widely-used pPXF \citep{ppxf} integrated in BADASS, adopting the Indo-U.S. Library of Coud\'e Feed Stellar Spectra \citep{Valdes2004} with an intrinsic model spectral broadening of FWHM $\sim$ 1\AA. The stellar velocity and velocity dispersion were free to vary in the fitting.
For emission lines, the broad \ha, \hb, and \hg\ emission lines exhibit double-peaked profiles and were fit with two Gaussian components with velocity and velocity dispersion free to vary in the ranges of [$-$3000, 3000] \kms\ and [500, 5000] \kms, respectively. Another two Gaussian components were adopted to fit the core (tracing the systemic gas) and blueshifted components (tracing the outflowing gas) of all narrow emission lines (\ha, \hb, \oiiiab, \nii). Here the term ``narrow emission lines'' is adopted solely to distinguish them from the double-peaked broad Balmer lines of the quasar, and does not indicate that they are all physically narrow. The kinematics (i.e., velocity and velocity dispersion) of each corresponding Gaussian component in all these lines are tied together, which are free to vary in the ranges of [$-$3000, 3000] \kms\ and [500, 5000] \kms, respectively. The flux ratio of the \oiiitext\ and \niitext\ doublets are fixed at their theoretical values, 1:2.98 and 1:3, respectively \citep{Osterbrock2006}. In addition, one Gaussian component was used to fit the He I emission line. Finally, all fluxes obtained from the best-fit models are corrected for aperture loss determined from the curve of growth analysis results from the NIRSpec/IFU data of the standard star TYC 4433-1800-1 (proposal ID: 1128) over the same wavelength range.

We adopt the best-fit stellar velocity to determine the systemic redshift of our object, which gives $z_{\rm sys} = 6.1831\pm{0.0001}$. The best-fit \sig\ is 288$\pm{13}$ \kms.
To better estimate the uncertainty of it, we also tested two other treatments for the power-law continuum in the fitting: (i) one with power-law index fixed to $-$1.7, a typical value for low-$z$ quasars \citep{Selsing2016} and free amplitude, and (ii) one with free power-law index but amplitude at 5050--5150 \AA\ matched to that from the best-fit quasar continuum flux from GALFIT. Both of these approaches gave larger \sig\ of $\sim$333$\pm{6}$ \kms\ and 341$\pm{7}$ \kms, respectively. We thus use the maximum difference between these two values and the best-fit value above as the uncertainty for the \sig, which gives 288$\pm{60}$ \kms. This is adopted as the fiducial value in the rest of the paper. This measurement is independent of the choice of aperture size as the stellar absorption lines are spatially unresolved in our IFU data. 
As a caveat, the \sig\ measured here is dominated by those of A-type and F-type stars and may not represent the kinematics of stellar populations with later spectral types.

\subsection{Comparison with the FS Spectrum}
\label{sec:32}

In \citet{Onoue2025}, they reported a lower \sig\ (a 2-$\sigma$ upper limit of \sig\ $<$ 190 \kms) for J1512$+$4422 based on the FS spectrum extracted from a 0\farcs6$\times$0\farcs2 aperture with about $\sim$2.7$\times$ worse spectral resolution. The absorption lines are also spatially unresolved within the FS data \citep{Phillips2025}.
A comparison of the IFU and FS spectra zoomed into the regions near the stellar Balmer absorption lines and the corresponding best-fit host galaxy models are shown in Fig. \ref{fig:comparespec} of Appendix \ref{append2}. In general, our IFU spectrum resembles the FS spectrum. Nevertheless, the Balmer absorption lines of the IFU spectrum are broader than those of the FS spectrum. Specifically, the FWHM of the \hd\ absorption, measured by fitting a Gaussian profile directly to the IFU spectrum, is 1.7$\times$ larger than that of the FS spectrum. This is the primary origin for the difference in \sig\ measured from the two spectra. In addition, the best-fit stellar continuum of the IFU spectrum is also slightly bluer than that of the FS spectrum (right panel of Fig. \ref{fig:comparespec}). 

In the fitting of the FS spectrum, they fixed the power-law index to $-$1.7 in the initial fit and finalized it to $\sim$$-$2.23 where they fit the host galaxy-subtracted spectrum obtained from the initial fit. They then tried to match the total quasar flux (both power-law and emission lines) from the FS spectrum to that from the JWST/NIRCam F356W image, but found that they had to scale up the quasar flux from the F356W image by 0.2 dex to keep the equivalent widths of Balmer absorption lines within the reasonable ranges of the stellar population models. Alternatively, they also tested flatter power-law indices but concluded that it would lead to a stellar continuum too blue to be reproduced by galaxy spectral energy distribution (SED) models. Moreover, broad \hb\ and \hg\ emission lines are modeled with a single Gaussian component in their fitting whereas our spectrum resolves the double-peak features of the broad \hb\ and tentatively so for the broad \hg, so these lines are modeled with two Gaussian components in our case.
Finally, as shown in our Fig. \ref{fig:comparespec} and the Extended Fig. 6 of \citet{Onoue2025}, the best-fit model cannot explain the red wing of \hg\ absorption line in the FS spectrum, which could again affect their best-fit \sig.

Our \sig\ is based on the higher spectral resolution IFU data alone and the contribution of quasar emission is determined from the image decomposition of the IFU data itself, which should suffer much less from the types of uncertainties described above. Nevertheless, our spectrum does have a $\sim$35\% lower continuum S/N per \AA\ near the absorption lines than that of the FS spectrum, which could make our \sig\ more uncertain. Moreover, while the Balmer absorption lines are spatially unresolved in both the IFU and FS data \citep{Phillips2025}, the deeper FS spectrum with a larger aperture could still have more contribution from the potentially spatially-extended, weak absorption lines and thus lead to a different \sig. 
Finally, we have also fit the FS spectrum of J1512$+$4422 with BADASS following the same parameter settings in \citet{Onoue2025}. It yields a \sig\ consistent with that reported in \citet{Onoue2025}, indicating that the choice of fitting software does not cause the difference between the IFU-based and FS spectrum-based results. In this paper, we use the \sig\ derived from our IFU data as the fiducial value for J1512$+$4422.

The other object in \citet{Onoue2025}, J2236$+$0032, does not exhibit the inconsistency in the decompositions of the quasar and host galaxy contribution between the FS spectrum and imaging for J1512$+$4422 as described earlier, and the reported absorption line width of it (270$\pm{60}$ \kms) is much larger than the spectral resolution.

As mentioned earlier, clear double-peaked broad \ha\ and \hb\ emission lines are seen in our IFU spectrum, which confirms J1512$+$4422 as the highest-redshift AGN with double-peaked broad Balmer emission lines, which was first reported in \citet{Onoue2025}. The double-peaked profiles of the \ha\ and \hb\ lines in the IFU spectrum resemble that of the \ha\ line as observed in the FS spectrum. However, no double-peaked profile is visible in the broad \hb\ of the FS spectrum, likely due to the lower spectral resolution.
Such double-peaked broad emission lines have been observed in lower-redshifts quasars/AGN, and many studies argue that they represent the emission from the relativistic Keplerian motion of emitting gas in a geometrically thin and optically thick accretion disk. \citep[e.g.,][]{Eracleous1994, Eracleous2009, Strateva2003,Ward2025}. More details on fitting the double-peaked broad \ha\ line of the FS spectrum with accretion disk models can be found in \citet{Onoue2025}.
A more comprehensive analysis of the double-peaked broad Balmer emission lines of the IFU spectrum will be presented in a future study.

\subsection{Extended Line Emission}
\label{sec:33}

We adopt \qtdfit\ \citep{q3dfit} to subtract the quasar PSF in the IFU data and uncover the extended line emission within the galaxy, following the same approach adopted by previous studies \cite[e.g.,][]{Wylezalek2022,Vayner2023b,Vayner2023c,VeilleuxLiu2023,Liu2024b, Liu2025a,Wolf2026}. Here we summarize the key steps of \qtdfit\ briefly. 
First, we construct a quasar template spectrum from an aperture ($r=$ 0\farcs{05}) centered on the brightest spaxel. For each spaxel in the IFU data cube, we then model the corresponding spectrum as a linear combination of a scaled quasar template spectrum representing the quasar PSF contribution in that spaxel ($I^n_{\rm quasar}$), a featureless monotonic polynomial representing the host-galaxy continuum emission ($I^n_{\rm starlight, exp.\;model}$), and a set of host-galaxy emission lines modeled with up to two components with Gaussian profiles ($I^n_{\rm emission}$). For the emission line, a second Gaussian component is kept only if it has peak S/N $>$ 3. In this analysis, we adopt no stellar population models in the fitting, since we only focus on the extended emission lines and see no evidence of spatially-extended stellar absorption lines.

As shown in Fig. \ref{fig:o3map}, the extended line emission in each spaxel can be modeled by a single Gaussian profile (c1 components) in general, except for several spaxels where a secondary Gaussian profile (c2 components) is needed (i.e., the c2 component has a peak S/N $>$ 3). Five example spectra from representative spaxels, along with their best-fit models of extended emission and subtracted quasar PSF contribution, are shown in Fig. \ref{fig:o3example}. The locations of these spaxels are denoted in Fig. \ref{fig:o3map}. The extended line emission mainly consists of two parts in terms of their kinematic nature: The \textit{Systemic Gas} includes the southern and northwestern parts of the c1 components with centroid velocities \vwu\ close to the systemic velocity ($<$50 \kms) and narrow line widths ($\sigma$ $<$ 100 \kms) in general. It traces the kinematically quiescent gas within the system. The \textit{Outflowing Gas} includes the remaining regions of the c1 components and the c2 components, which are all blueshifted (\vwu\ $\sim$ $-$20 -- $-$1040 \kms; median: $\sim$$-$174 \kms) and have much broader line widths ($\sigma$ $\sim$ 100 -- 700 \kms; median: $\sim$156 \kms) with respect to the \textit{Systemic Gas}. It traces a galaxy-scale outflow in this system. The approximate locations of
the c1 components belong to the \textit{Outflowing Gas} are indicated by the red ellipses in Fig. \ref{fig:o3map}. In this paper, to derive the outflow energetics, we define the \textit{Outflowing Gas} as those traced by Gaussian components (c1 or c2) with either (i) \vwu\ $<$ 0 \kms\ and $\sigma$ $>$ 100 \kms\ or (ii) \vwu\ $<$ $-$100 \kms. These thresholds for \vwu\ and $\sigma$ are chosen to be approximately the maximum values observed for the \textit{Systemic Gas} in our object, and are also more than 1-$\sigma$ away from the corresponding median values of the \textit{Systemic Gas}.

Nevertheless, it should be noted that we cannot formally rule out other interpretations for the nature of the \textit{Outflowing Gas}. For example, it could still come from the host galaxy itself but with much more turbulent motions as reflected by its larger line width than the \textit{Systemic Gas}. However, this scenario does not automatically explain the observed coherent blueshift of the line emission. Part of the broad and blueshifted line emission could potentially trace a combination of outflowing and rotating gas, although the current data are not be deep enough to disentangle the two. 
Moreover, the entire line-emitting nebulae may still trace the rotating gas, although the gas kinematics observed is apparently different from those of normal rotating disks where the line emission would shift from blueshift to redshift more gradually.

\begin{figure*}
    \centering
    \begin{minipage}{\textwidth}
    \includegraphics[width=\linewidth]{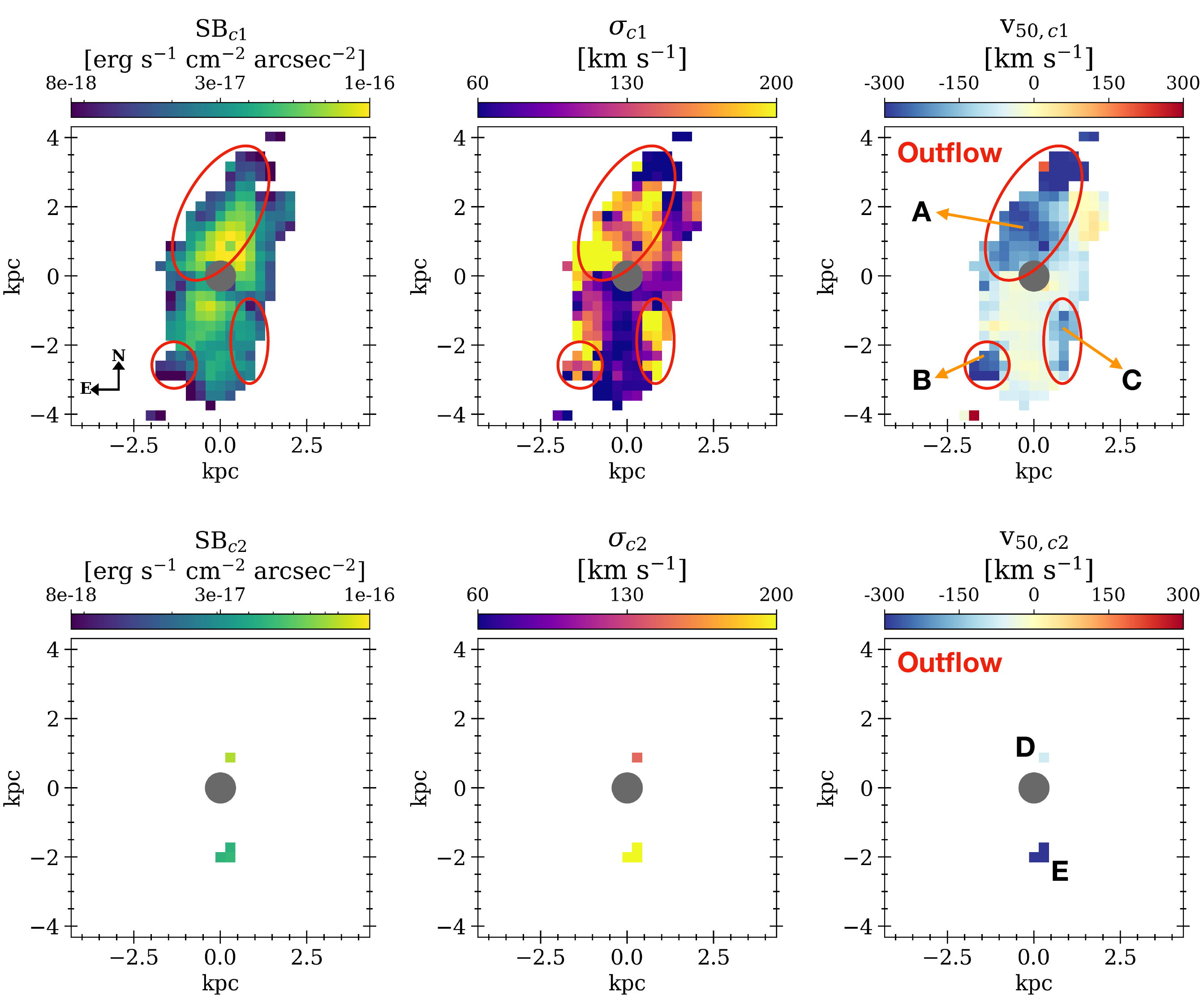}
        \end{minipage}
    \caption{Flux, velocity dispersion $\sigma$, and centroid velocity \vwu\ maps of the \oiii-emitting gas in our object J1512$+$4422, after the quasar PSF has been subtracted. The top row is for the narrower Gaussian components (c1 components) and the bottom row is for the broader Gaussian components (c2 components). Only components with peak flux density above 3$\sigma$ and passing visual inspection are kept and shown in these figures. The approximate locations of the c1 components associated with the \textit{Outflowing Gas} are indicated by the red ellipses. All c2 components belong to the \textit{Outflowing Gas}. See Sec. \ref{sec:33} for more details. The labels A--E denote the spaxels whose spectra and best-fit models are shown in Fig. \ref{fig:o3example}. The location of the quasar is indicated by the gray circle. The PSF size is $\sim$0.8 kpc.}
    \label{fig:o3map}
\end{figure*}

\begin{figure*}
    \centering
    \begin{minipage}{0.32\textwidth}
     \includegraphics[width=\linewidth]{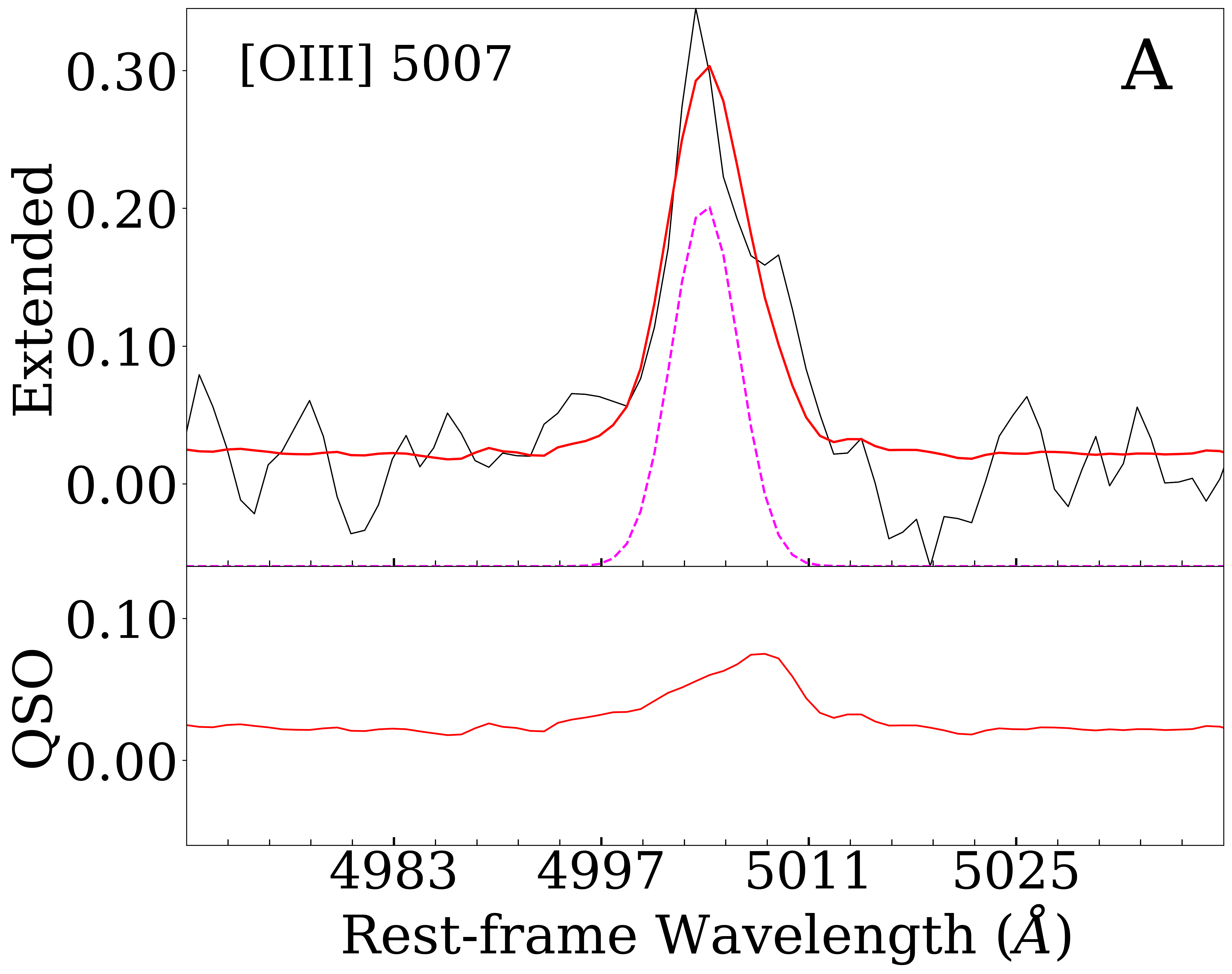}   
    \end{minipage}
    \begin{minipage}{0.32\textwidth}
     \includegraphics[width=\linewidth]{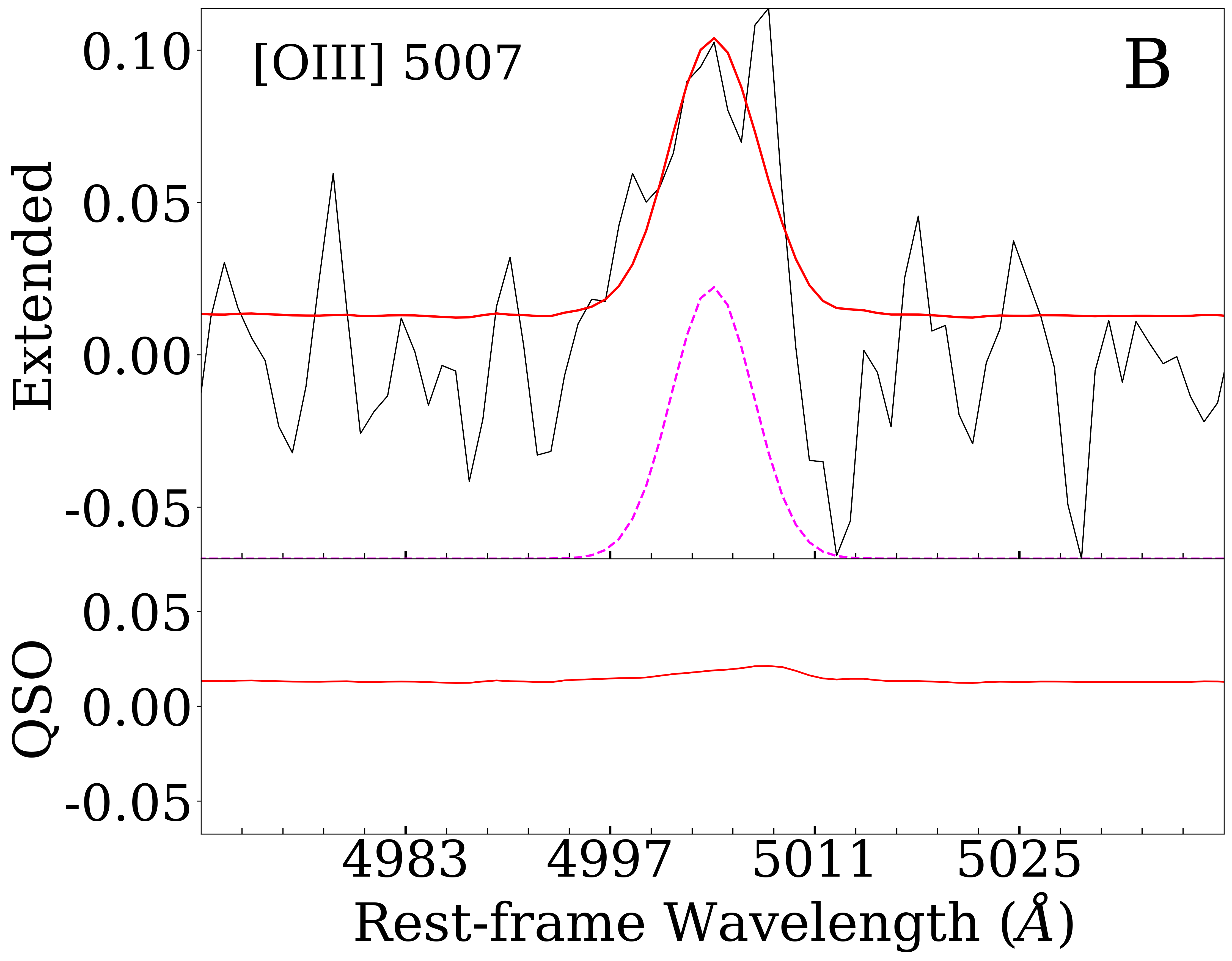}   
    \end{minipage}
        \begin{minipage}{0.32\textwidth}
     \includegraphics[width=\linewidth]{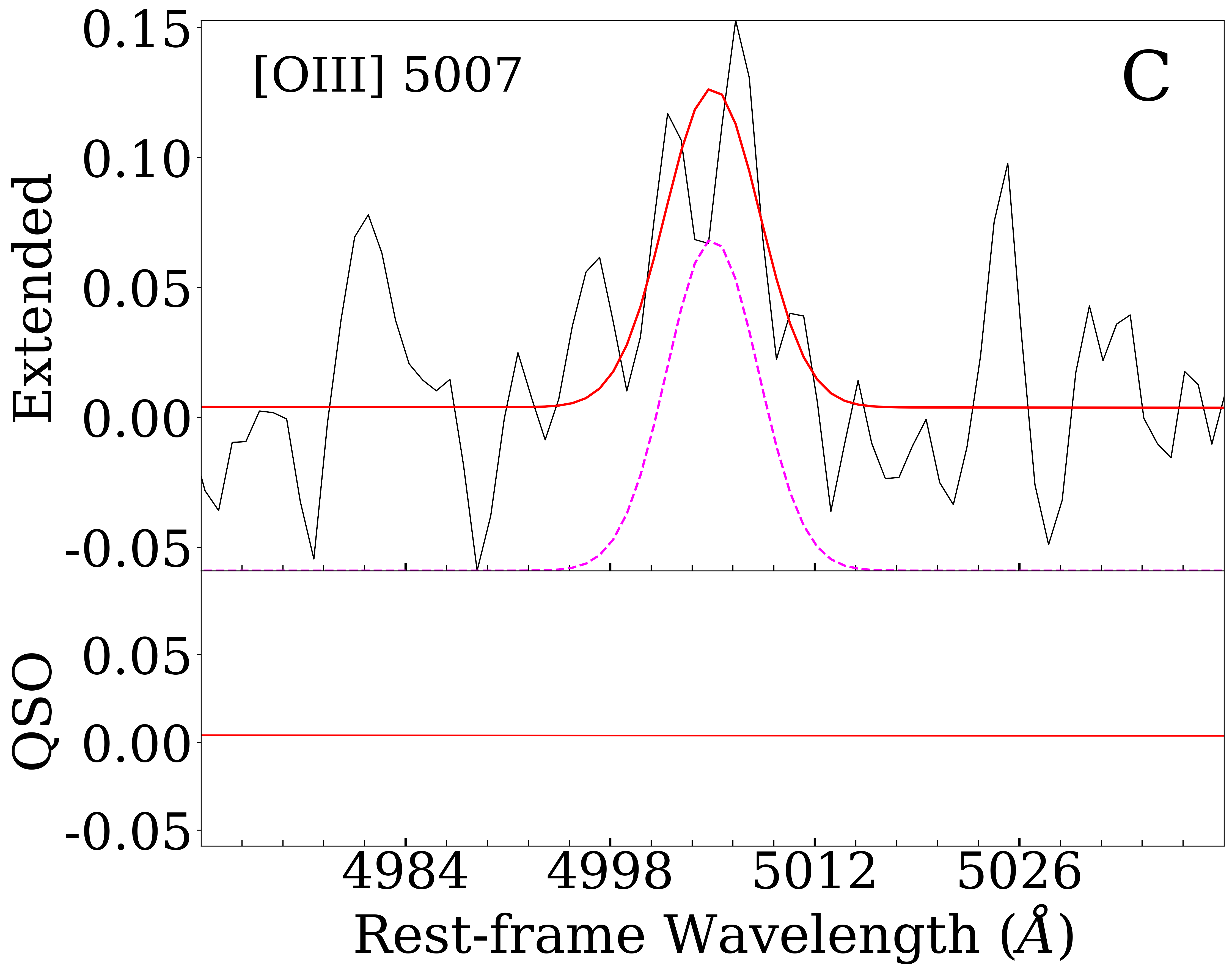}   
    \end{minipage}
    
        \begin{minipage}{0.32\textwidth}
     \includegraphics[width=\linewidth]{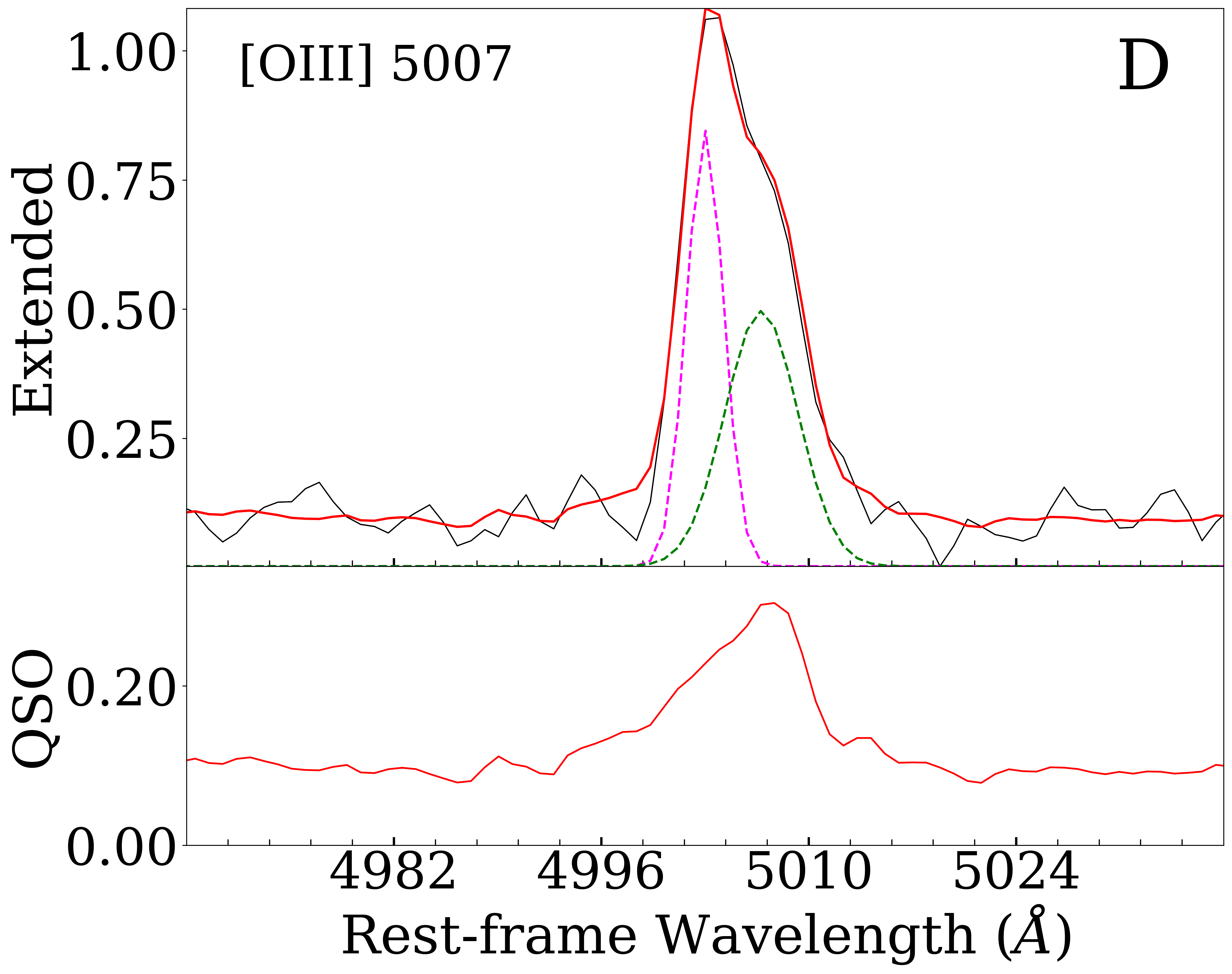}   
    \end{minipage}
        \begin{minipage}{0.32\textwidth}
     \includegraphics[width=\linewidth]{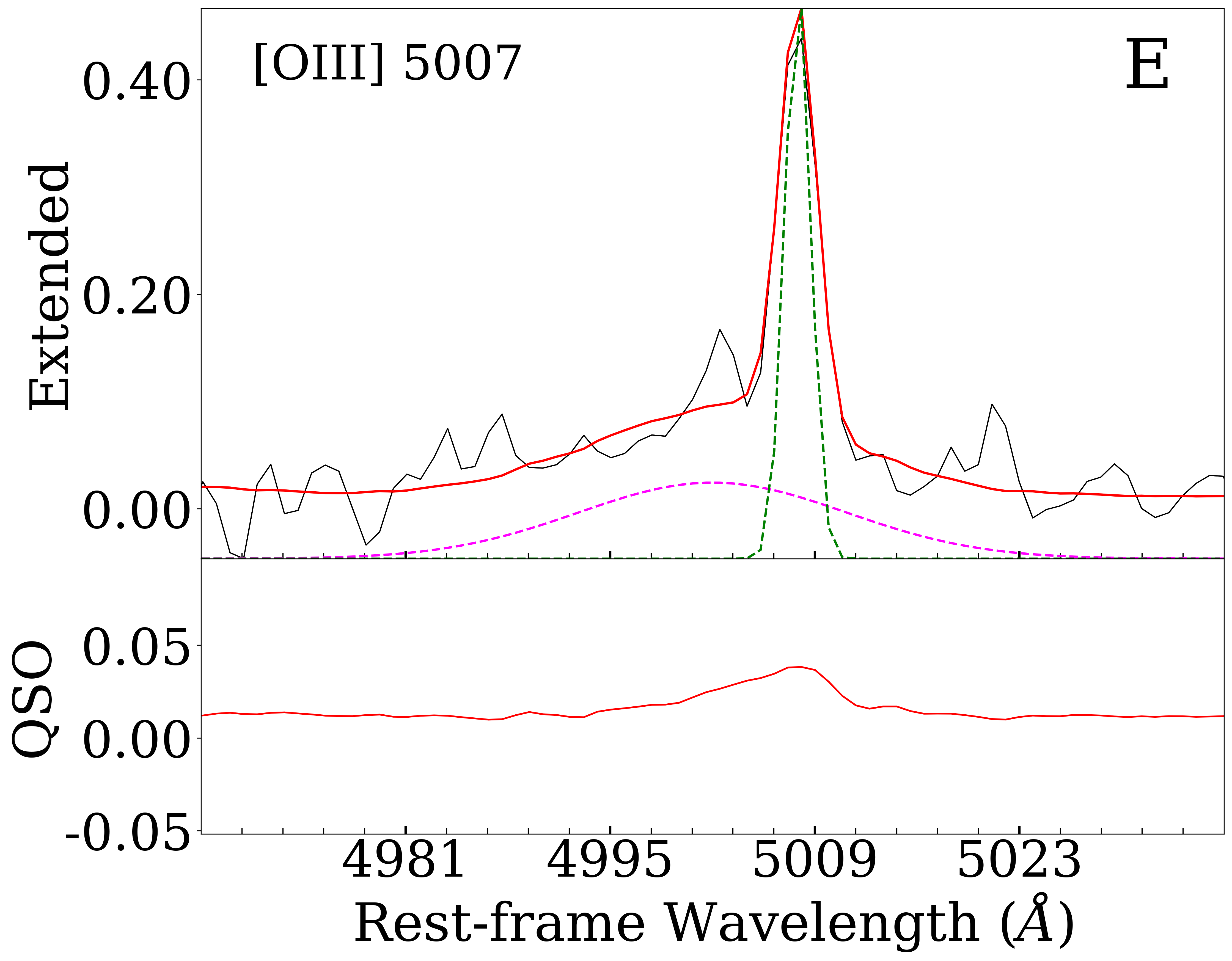}   
    \end{minipage}
    \caption{\oiiitext\ spectra from the 5 representative spaxels (A--E) shown in Fig. \ref{fig:o3map}.
    In each panel: \textbf{Top:} Spectrum of the extended emission (black; i.e., with the quasar PSF contribution subtracted), the overall best-fit model in red, and the individual Gaussian components of best-fit emission line model in magenta and green dashed lines. \textbf{Bottom:} The best-fit model spectrum of the quasar PSF contribution in the spaxel. In other words, the spectrum in the top is obtained by subtracting this model spectrum from the original spectrum. The y axes are in arbitrary flux density units. See Sec. \ref{sec:33} for more details on the subtraction of quasar PSF contribution in each spectrum.}
    \label{fig:o3example}
\end{figure*}

\section{A Nearly One Billion Solar Mass SMBH on the Local $M_{\rm BH}$ -- $\sigma_{\rm \ast}$ Relation}\label{4}

\subsection{Quasar Properties}
\label{41}

The bolometric luminosity of our object is derived adopting the 5100\AA\ continuum luminosity (\Lwu) of the quasar with 
 a bolometric correction factor of 9.26 \citep{Richards2006}, which gives log($L_{\mathrm{bol}}$/erg s$^{-1}$) $\sim$ 45.1.
The \hb-based and \ha-based BH mass are derived following \citet{Vestergaard2006} and \citet{GreeneHo2005}, respectively, which are:

\begin{equation}
\begin{split}
{\rm log}\left(\frac{M_{\rm BH, H\beta}}{M_\odot}\right) = {\rm log}\Bigg\{\left[\frac{{\rm FWHM}({\rm H\beta})}{1000\ {\rm km\ s}^{-1}}\right]^2 \\ \times\left[\frac{\lambda L_\lambda(5100 )}{10^{44}\ {\rm erg\ s}^{-1}}\right]^{0.5}
\Bigg\}
+ (6.91 \pm{0.02})
\end{split}
\label{eq:BH}
\end{equation}

\begin{equation}
\begin{split}
{\rm log}\left(\frac{M_{\rm BH, H\alpha}}{M_\odot}\right) = {\rm log}\Bigg\{(2.0^{+0.4}_{-0.3}\times10^6)\\
\times\left[\frac{{\rm FWHM}({\rm H\alpha})}{1000\ {\rm km\ s}^{-1}}\right]^{2.06\pm{0.06}}\\
\times\left[\frac{L({\rm H\alpha})}{10^{42}\ {\rm erg\ s}^{-1}}\right]^{0.55\pm{0.02}} \bigg\}
\end{split}
\label{eq:BHHa}
\end{equation}

Here FWHM(\hb) and FWHM(\ha) are calculated for the entire line profile of the broad \hb\ and \ha\ emission lines, respectively. The Eddington ratios (\edd) are then derived following $L_{\rm Edd} = 1.26\times10^{38} (M_{\rm BH}/M_\odot)$ erg s$^{-1}$. J1512$+$4422 has a \mbh\ of $\sim$6.2$\times$10$^{8}$ \msun\ and \edd\ of $\sim$0.016 based on \hb, and a \mbh\ of $\sim$8.8$\times$10$^{8}$ \msun\ and \edd\ of $\sim$0.012 based on \ha, respectively. The \hb-based and \ha-based results are consistent with each other considering the systematic uncertainties of them. These results are also listed in Table \ref{tab:quasar}. In \citet{Onoue2025}, they report similar \mbh\ and \edd\ (\hb: $\sim$4.8$\times$10$^{8}$ and 0.11; \ha: $\sim$1.3$\times$10$^{9}$ \msun and 0.03) based on the FS spectrum. 
Nevertheless, one caveat of the measurements above is that we assume Eqs. \ref{eq:BH} and \ref{eq:BHHa} are applicable to quasars with double-peaked \ha\ and \hb\ profiles following previous studies \cite[e.g.,][]{Wu2004,Onoue2025}, which may not hold \citep[e.g.,][]{Bian2007}: As mentioned earlier in Sec. \ref{sec:32}, double-peaked broad Balmer emission lines could represent the relativistic Keplerian motion of a geometrically thin and optically thick accretion disk rather than the broad line region (BLR).
In the remainder of this paper, we use our \ha-based values as the fiducial ones.

\subsection{Locations on the SMBH--Galaxy Scaling Relations}
\label{42}

As shown in the left panel of Fig. \ref{fig:msigma}, our new data suggest that J1512$+$4422 at $z \sim$ 6.2 already lies on the local $M_{\rm BH}$--$\sigma_\ast$ relation \citep{KormendyHo2013}, like the other object from \citet{Onoue2025}, J2236$+$0032, at $z \sim$ 6.4. 
These two objects also fall within the region occupied by z$<$1 quasars with post-starburst signatures from the SDSS Reverberation Mapping project \citep{Matsuoka2015}. Nevertheless, their SMBHs are still overmassive when compared to the local $M_{\rm BH}$--$M_{\ast}$ relation \citep{KormendyHo2013,Reines2015}. 
At $\sim$900 million years after the Big Bang, the dynamics and gravitational potential wells of (at least the inner regions of) these two quasar host galaxies are already as mature as those observed in the present-day universe. The tight $M_{\rm BH}$--$\sigma_\ast$ relation is likely being established at $z\sim6$ for such quasars. 

Previous studies have mainly proposed two mechanisms to explain the origin of this tight correlation: Some argue that it is a natural consequence of frequent mergers during the evolution of the systems \citep[e.g.,][]{Jahnke2011,KormendyHo2013}. The others suggest that AGN feedback plays a critical role in shaping it \citep[e.g.,][]{Costa2014b,king15}. For the merger scenario, some studies do suggest that such $z>$ 6 quasars could have gone through frequent mergers \citep[e.g.,][]{Li2007,Lupi2019}, although such scenario may only hold for very extreme cases. For example, we use \textit{TRINITY}, an empirical model connecting dark matter halos, galaxies, and SMBHs from $z=$ 0 -- 10 \citep{trinity,trinity4}, to estimate the merger rates for halos hosting quasars like J1512$+$4422. In a volume of (250 cMpc h$^{-1}$)$^3$, the average number of major mergers (mass ratio $>$ 1:10) from $z=$ 6--10 for $z=$ 6 halos above 10$^{12}$ \msun\ is $\sim$0.10, with only $\sim$1\% of the halos having two mergers and no halos having more than two.
For the feedback scenario, evidence of effective quasar feedback via outflows is indeed present in J1512$+$4422 and likely also in J2236$+$0032, as discussed in Sec. \ref{5} below. Overall, it is likely that both mechanisms help to establish the $M_{\rm BH}$--$\sigma_\ast$ in the Reionization Era together.

On the other hand, \citet{Onoue2025} reported the $M_{\rm \ast}$ of J1512$+$4422 and J2236+0032 ($M_{\rm \ast}/M_\odot$ $=$ 10.64$^{+0.04}_{-0.01}$ and 10.80$^{+0.03}_{-0.02}$, respectively), which were derived by fitting both the JWST NIRCam photometry and FS spectra of the host galaxy with stellar population models using \textit{BAGPIPES}. As shown in the right panel of Fig. \ref{fig:msigma}, their $M_{\rm \ast}$ are still slightly outside the dispersion of the local scaling relation, even though they are closer to the relation than the majority of $z>5$ quasars and AGN reported in the literature \citep[e.g.,][]{yue_eiger_2023,Maiolino2024}.
While in a rapid quenching phase, these two objects may still need to gain some stellar mass to land on the local relation. This could be achieved by mergers and/or rejuvenation (i.e., a new episode of star formation/starburst in the future).
Nevertheless, it should be noted that stellar masses may suffer from large uncertainties in quasar host galaxies, which can be significantly underestimated due to the fact that (i) the quasar PSF emission could be oversubtracted in imaging decomposition (ii) the mass of later-type stellar populations may not be accounted for \citep[i.e., outshining effect, e.g., ][]{Conroy2013}. For J1512$+$4422, the uncertainty could be even larger due to the limited wavelength coverage of photometry and spectroscopy adopted in the SED fitting \citep{Onoue2025}. Moreover, \citet{Ding2023,Ding2025ApJ,Silverman2025} report that many $z\sim6$ SHELLQs quasars, with one of them being J2236$+$0032, are already consistent with the local $M_{\rm BH}$-$M_\ast$ relation, after selection effects have been taken into account. Therefore, the $M_{\rm BH}$ and $M_\ast$ of J1512$+$4422 could also be consistent with the local relation, as the apparently smaller $M_\ast$ may be also caused by similar selection effects.

Altogether, within 1 billion years after the Big Bang, the discoveries of two quasars already located on the local $M_{\rm BH}$-$\sigma_\ast$ relation \citep[this work and][]{Onoue2025}, along with quasars whose properties are potentially consistent with the local $M_{\rm BH}$-$M_\ast$ relation \citep{Ding2023}, suggest that the formation and evolution of certain SMBHs and galaxies are coupled with each other very early on.

The gas velocity dispersion \siggas\ (86$\pm{9}$ \kms) derived from the core component of the narrow emission lines in J1512$+$4422 is much smaller than the \sig, but close to the \siggas\ (149$\pm{4}$ \kms) measured from the FS spectrum with lower spectral resolution. Similarly, J2236$+$0032 also exhibits much smaller \siggas\ (140$\pm{17}$ \kms) than \sig\ (270$\pm{60}$ \kms). One possible explanation is that the line emission is dominated by the brightest, quasar photoionized regions and/or \ion{H}{2} regions. The emission line width thus only reflects the internal kinematics of those ionized regions rather than the global kinematics of the galaxy.
This is supported by the lack of clear sign of rotation in the gas velocity field of J1512$+$4422. Instead, the ionized gas is either highly blueshifted, tracing a rapid outflow or close to the systemic velocity determined by the stellar absorption line.
These suggest that the gas velocity dispersion is not a good indicator of stellar velocity dispersion in high-$z$ quasars, at least for those like J1512$+$4422, and should be used with caution when investigating correlations like the $M_{\rm BH}$-$\sigma_\ast$ relation. Similar conclusions have also been reached by studies of lower-redshift quasar/AGN \citep[e.g.,][]{Greene2005,Remigio2025}.

\begin{deluxetable}{ll}
\tablecaption{Quasar and Host Galaxy Properties of J1512$+$4422 based on the IFU Data}
\tablehead{
\colhead{Property} & \colhead{Value}
}
\startdata
redshift (stellar)  &  $6.1831\pm{0.0001}$ \\
log($L_{\mathrm{5100}}$/erg s$^{-1}$) &  $44.131\pm{0.001}$ \\
log($L_{\mathrm{bol}}$/erg s$^{-1}$) & 45.1 \\
FWHM$_{\rm broad}$ (\kms) & $8048\pm{134}$ \\
log($M_{\mathrm{BH, H\beta}}$/\msun) & $8.79\pm{0.01}$ \\
$\lambda_{\rm Edd, H\beta}$ & $0.016\pm{0.001}$ \\
log($M_{\mathrm{BH, H\alpha}}$/\msun) & $8.94\pm{0.02}$  \\
$\lambda_{\rm Edd, H\alpha}$ & $0.011\pm{0.001}$ \\
\sig\ (\kms) &  $288\pm{60}$ \\
\siggas\ (\kms) & $86\pm{9}$ \\
\enddata
\tablecomments{FWHM$_{\rm broad}$: The overall FWHM of the broad Balmer lines. The uncertainties listed only include those from the spectral fitting.}
\label{tab:quasar}
\end{deluxetable}

\begin{deluxetable*}{c cccc ccc}[!htb]
\tablecaption{Outflow Properties\label{tab:energetics}}
\tablehead{
& \colhead{$L_{\mathrm{[O III]}}$} &
 \colhead{$V_{\mathrm{out}}$}  &
   \colhead{$R_{\mathrm{out}}$}  & 
   \colhead{$\dot{M}_{\mathrm{out}}$} & \colhead{$\dot{p}_{\mathrm{out}}$}  & \colhead{$\dot{E}_{\mathrm{out}}$} \\
  & \colhead{[$10^{42}$ erg s$^{-1}$]} &
 \colhead{[\kms]}  &
   \colhead{[kpc]}   & \colhead{[\msunyr]} & \colhead{[$\times10^{34}$ dynes]}  & \colhead{[$\times10^{41}$ erg s$^{-1}$]} \\
& \colhead{(1)} & \colhead{(2)} & \colhead{(3)} & \colhead{(4)} & 
\colhead{(5)} & 
\colhead{(6)}
}
\startdata 
{Nuclear} & $6.6\pm{0.3}$ & $478\pm{14}$ & $<$0.8 & $92.6^{+92.6}_{-74.1}$ & $28.1^{+28.1}_{-22.5}$ & $67.1^{+67.1}_{-53.7}$  \\
Extended & 6.6$\pm{1.6}$ & $352\pm{75}$ & 3.2 & $5.1^{+5.1}_{-4.1}$ & $1.3^{+1.3}_{-1.0}$ & $4.1^{+4.1}_{-3.3}$ 
\enddata
\tablecomments{Outflow properties of J1512$+$4422 for the spatially unresolved, nuclear outflow (first row) and spatially resolved, extended outflow (bottom row). From left to right, the columns are: (1) \oiiitext\ luminosity; (2) outflow velocity (defined as $|v_{50}|+\sigma$). For the extended outflow, this is the median value of all spaxels with outflowing gas; (3) radial distance; (4) mass outflow rate; (5) momentum outflow rate; and (6) kinetic energy outflow rate. Here the uncertainties for columns (4)--(6) correspond to those caused by the electron density range of 100--1000 cm$^{-3}$.}
\end{deluxetable*}

\begin{figure*}[!htb]
    \centering
    \begin{minipage}{0.49\textwidth}
    \includegraphics[width=1\linewidth]{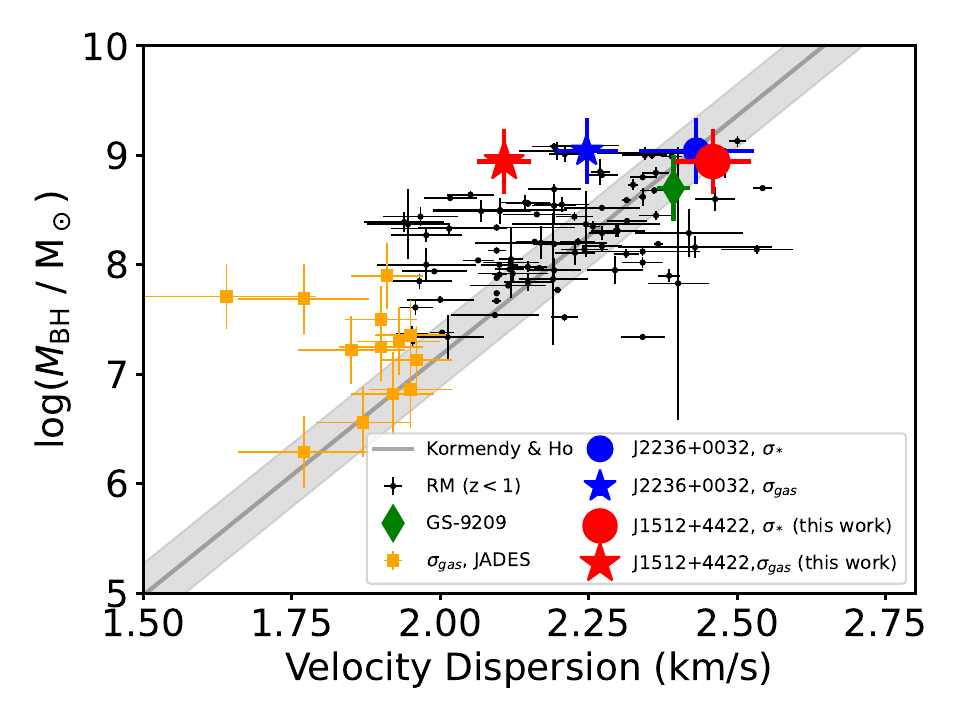}
    \end{minipage}    
    \begin{minipage}{0.48\textwidth}
    \includegraphics[width=1\linewidth]{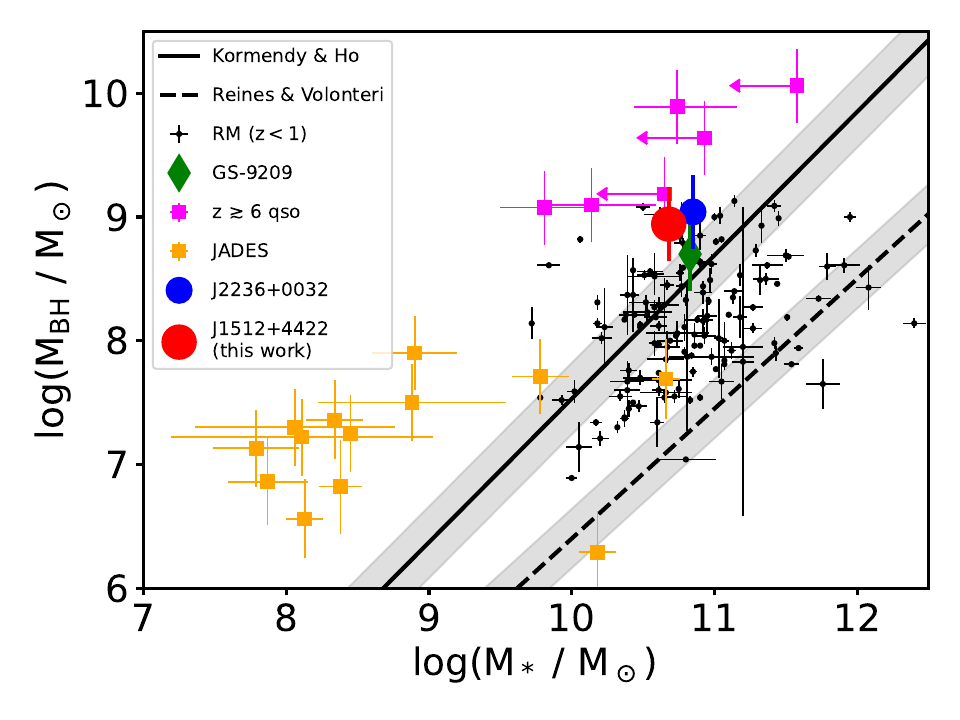}
    \end{minipage}
    \caption{BH masses versus velocity dispersions (left) and stellar masses (right) for our objects and other quasars and AGN. The error bars of the BH masses of J1512$+$4422, J2236$+$0032 and GS-9209 represent the typical systematic uncertainties \citep[0.3 dex; e.g.,][]{Maiolino2024}. \textbf{Left:} \sig: our object J1512$+$4422 (red circle) based on the IFU data and J2236$+$0032 from \citet{Onoue2025}. The $z\sim4.7$ quiescent galaxy GS-9209 \citep{Carnall2023} and $z<1$ RM quasars \citep{Matsuoka2015} are denoted by the green diamond and black points, respectively. \siggas: J1512$+$4422 and J2236$+$0032 are denoted by the red and blue stars, respectively. The orange points represent high-$z$ AGN from the JADES program \citep{Maiolino2024}. The gray solid line and associated shaded band represent the local relation and its dispersion \citep{KormendyHo2013}. \textbf{Right:} The data points are denoted with the same symbols shown in the left panel. Additionally, the magenta squares denote luminous $z\gtrsim6$ quasars from \citet{yue_eiger_2023}. Shown in the black solid and dotted-dashed lines are the local relations from \citet{KormendyHo2013} and \citet{Reines2015}, respectively. The gray bands indicate the corresponding dispersions.}
    \label{fig:msigma}
\end{figure*}

\section{A Fast Outflow Quenching the Galaxy}
\label{5}

Quasar/AGN feedback via outflows is expected to be an important mechanism for quenching \citep[e.g.,][]{DiMatteo2005,Sijacki2009,zubo12,fabi12,Costa2014a,schaye_eagle_2015,Veilleux20,Liu2020,Bennett2024, Husko2025,Liu2026a}. Nevertheless, within the first billion years of the cosmic history, direct observations of how such feedback-induced quenching happens are still lacking. In the following, we demonstrate that the quasar feedback via outflow is the most promising mechanism leading to the rapid quenching of J1512$+$4422.

\subsection{Outflow Time Scale}
\label{51}

As described in Sec. \ref{sec:33}, we detect a fast outflow in J1512$+$4422 that likely extends on galaxy scale. Here we estimate how long this outflow has traveled, using 

\begin{equation}
t_{\rm out} = R_{\rm out} / v_{\rm out}
\label{eq:time}
\end{equation}

Here $R_{\rm out}$ $\simeq$ 3.2 kpc is estimated as the maximum radial distance of the extended outflow in projection. We define outflow velocity as $v_{\rm out}=|v_{50}|+\sigma$. As adopted in previous studies \citep[e.g.,][]{Vayner2023c,Liu2024b}, this formalism of outflow velocity is adopted to account for the inclination correction needed to recover the true outflow velocity in the 3D space. The line width is included as the $\sigma$ term and encodes both the outflow velocity out of the line of sight and the turbulent motion of the outflowing gas.
In the Eq. \ref{eq:time} above, $v_{\rm out}$ is estimated as the median value across all spaxels with outflowing gas ($352\pm{75}$ \kms). The corresponding $t_{\rm out}$ is then $\sim$ 9 Myr. 

As stated earlier, \citet{Onoue2025} suggests that the sSFR of J1512$+$4422 drops below 0.2 Gyr$^{-1}$ within the last 10 Myr. This time scale is comparable to the time scale of the outflow derived above, which supports the scenario that the outflow plays a major role in the current quenching process.

\subsection{Outflow Energetics}

\subsubsection{Nuclear Outflow}
\label{521}

In the nuclear spectrum (Fig. \ref{fig:nuclear}), the \oiii\ emission line has a prominent blueshifted and broad component, tracing a fast outflow in our object. 
The luminosity of the broad \oiii\ emission line component from the best-fit in Fig. \ref{fig:nuclear} is adopted to derive the mass of the outflowing gas following \citet{Can2012, Vei2020}:

 \begin{eqnarray}
M_{\rm ionized} & = & 5.3 \times 10^8~\frac{C_e L_{44}([\mathrm{O~III}])}{n_{e,2} 10^{\mathrm{[O/H]}}} M_\odot,
\label{eq:M_ionized}
\end{eqnarray}
where $L_{44}([\mathrm{O~III}])$ is the dust-extinction corrected luminosity of \oiiitext\ normalized to 10$^{44}$ erg s$^{-1}$. Here we assume case B conditions with an electron temperature $T \sim 10^4$ K \citep{Osterbrock2006}. We use the overall non-BLR \ha/\hb\ ratio ($\sim$6.0) to estimate the dust extinction of the outflowing gas, adopting the extinction law from \citet{CCM1989} and an intrinsic \ha/\hb\ ratio of 3.1 \citep{Osterbrock2006}. This gives E(B$-$V) $\simeq$ 0.7 and a correction factor of $\sim$8.4 for \oiiitext\ luminosity. Such dust extinction is a bit high but still within the ranges seen in $z\sim6$ quasar host galaxies \citep[e.g.,][]{Decarli2024}. The dust extinction derived from the \ha/\hb\ ratio of the outflow component alone is even higher, although it suffers from much larger uncertainties and thus not adopted in our calculation. The quantity $n_{e,2}$ is the average electron density, normalized to 100 cm$^{-3}$. It is set to 2, a value widely adopted in previous studies of quasar-driven outflows and comparable to the observed values when direct measurements are available \citep[e.g.,][]{Liu2013b,Harrison2014}. A range of $\sim$1--10 for $n_{e,2}$ is as reported in the literature \citep[e.g.,][]{Harrison2018}, which is adopted to estimate the uncertainties of outflow dynamics. 
The quantity $C_e \equiv \langle n_e \rangle^2 / \langle n_e^2 \rangle $ is the electron density clumping factor, which can be assumed to be of order unity on a cloud-by-cloud basis (i.e., each gas cloud has uniform density). The oxygen-to-hydrogen abundance ratio relative to the solar value, [O/H] is assumed to be 0 (i.e., solar oxygen abundance) in our calculation. 
Finally, we obtain a gas mass of $(1.5\pm{0.1})\times10^8$ \msun\ for the outflow.

To estimate the mass, momentum, and kinetic energy outflow rates, we follow the same approach adopted in \citet{Liu2024b}, namely:

\begin{eqnarray}
\dot{M}_{\rm nuc} & = & M_{\rm nuc}(v_{\rm nuc}/R_{\rm nuc}) \\
\dot{p}_{\rm nuc} & = & \dot{m}_{\rm nuc}v_{\rm nuc} \\
\dot{E}_{\rm nuc} & = & \frac{1}{2}~\dot{m}_{\rm nuc}(v_{\rm nuc})^2 
\end{eqnarray}

Here the size of the outflow $R_{\rm nuc}$ is assumed to be half of the PSF, which is $\sim$0.8 kpc. The outflow velocity is again measured as $v_{\rm nuc} = |v_{50}| + \sigma$, which gives $\sim$478$\pm{14}$ \kms.
The final results are listed in Table \ref{tab:energetics}. Our measured outflow velocity is smaller than that (721$\pm{39}$ \kms) measured from the FS spectrum with a lower spectral resolution and a much larger extraction aperture \citep[0\farcs6$\times$0\farcs2,][]{Onoue2025}.

\subsubsection{Extended Outflow}

Following \citet{Liu2024b}, we then calculate the mass, momentum and kinetic energy outflow rates of the extended outflow by integrating over all spaxels tracing the outflowing gas (see Sec. \ref{sec:33} for details) in the PSF-subtracted data cube, following:  

\begin{eqnarray}
\label{eq:Mdot}
\dot{M} & = & \Sigma~\dot{m_i} = \Sigma~m_i~(v_i/R_i) \\
\label{eq:pdot}
\dot{p} & = & \Sigma~\dot{m_i}~v_i \\
\label{eq:Edot}
\dot{E} & = & \frac{1}{2}~\Sigma~\dot{m_i}(v_i)^2 
\end{eqnarray}

Here the ionized gas mass of each spaxel, m$_i$, is calculated following Eq. \ref{eq:M_ionized}, except that we apply no dust extinction correction to the \oiiitext\ luminosity since we have no robust detection of \hb\ for the extended emission.  The outflow velocity of each spaxel, $v_i$, is defined in Section \ref{51} and has a maximum of $\sim$1740 \kms\ and a median of 352 \kms. $R_i$ is the projected distance to the quasar of each spaxel. The final results are listed in Table \ref{tab:energetics}. The slit of the FS spectrum from \citet{Onoue2025} does not cover the majority of the extended outflow, so we make no comparison between the two results here.

\subsubsection{Energy Source and Impact of the Outflows}

\begin{figure*}[!htb]
    \centering
    \begin{minipage}{0.49\textwidth}
    \includegraphics[width=\linewidth]{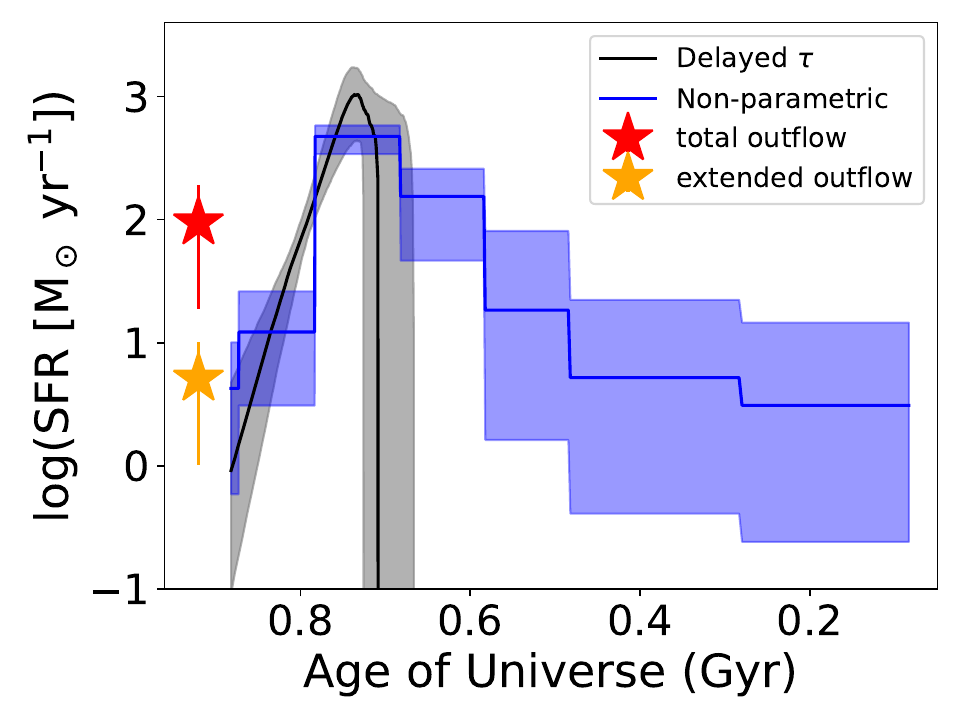}
    \end{minipage}
    \begin{minipage}{0.49\textwidth}
    \includegraphics[width=\linewidth]{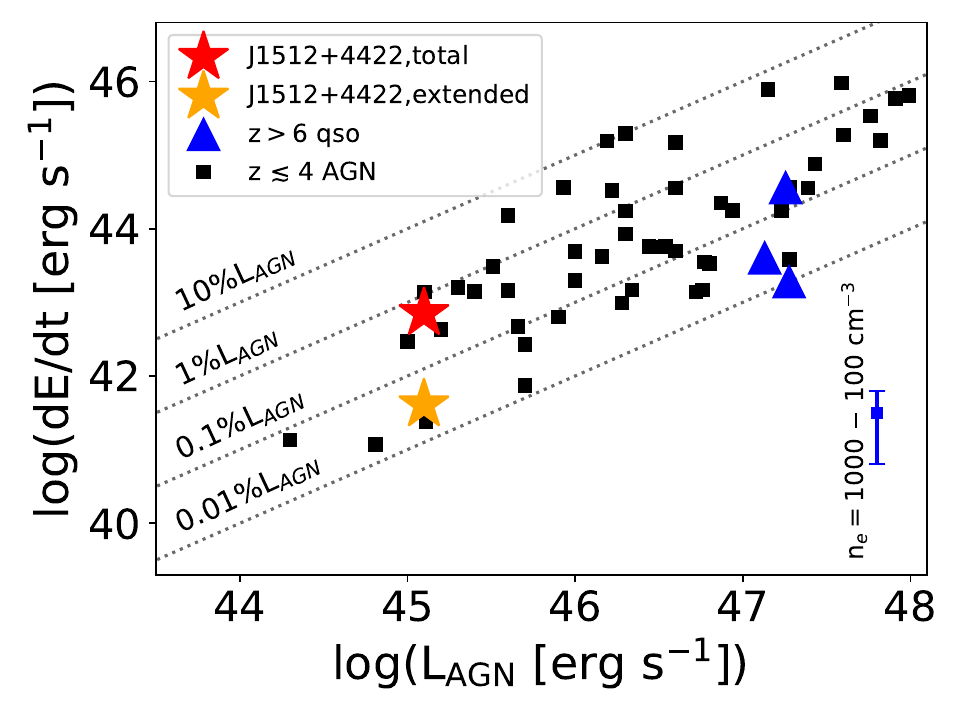}
    \end{minipage}
    \caption{Outflow energetics. \textbf{Left:} Stars: the mass outflow rates of the extended (orange) and total outflow (the nuclear and extended outflows combined; red). The error bars reflect the uncertainties corresponding to the $n_e$ range of 100--1000 cm$^{-3}$. Solid lines: the delayed-$\tau$ SFH (black) and non-parametric SFH (blue) obtained in \citet{Onoue2025}. The associated shaded regions indicate 16th--84th percentile intervals. \textbf{Right:} Kinetic energy outflow rate of the extended (orange) and total (red) outflow in our object J1512$+$4422 as a function of bolometric AGN luminosity. Also plotted are examples of spatially-resolved galaxy-scale outflows in other $z>6$ quasars revealed by NIRSpec/IFU data \citep[blue triangles; ][]{Marshall2023, Liu2024b} and other examples of galaxy-scale AGN-driven outflows at $z<4$ compiled by \citet{Fiore2017}. The uncertainties caused by electron density ($n_{e}=\ $1000 -- 100 cm$^{-3}$) is indicated by the blue bar at the bottom right corner.}
    \label{fig:energetics}
\end{figure*}

As shown in the left panel of Fig. \ref{fig:energetics}, the SFR of J1512$+$4422 within the last 10 Myr, when assuming either delayed-$\tau$ star formation history (SFH; 0.9$^{+3.8}_{-0.8}$ \msunyr) or non-parametric SFH (4.3$^{+5.8}_{-3.7}$ \msunyr) as derived in \citep{Onoue2025}, is smaller than the total mass outflow rate (97.7$^{+97.7}_{-78.2}$ \msunyr). Here the SFHs are retrieved from \citep{Onoue2025} which models the FS spectrum and NIRCam F150W and F356W images of the host galaxy simultaneously with \textit{BAGPIPES}. The best-fit SFHs remain similar if we instead adopt the IFU spectrum of the host galaxy in the modeling.
The current star formation is thus not able to drive the observed outflow. While J1512$+$4422 experiences a starburst about $\sim$150 Myr ago with a peak SFR of 1400$^{+520}_{-430}$ \msunyr, it is much earlier than the launching time ($\lesssim$9 Myr ago) of the current outflow and thus not responsible for it. On the contrary, with a kinetic energy outflow rate to quasar bolometric luminosity ratio of $0.6^{+0.6}_{-0.5}$\%, the quasar can easily drive the observed outflow and should be the main power source of it.

The total momentum outflow rate is $2.9^{+2.9}_{-2.4}\times10^{35}$ dynes, which is $6.7^{+6.7}_{-5.4}\times$ the momentum flux from the quasar radiation. This significant momentum boost suggests that the outflow could be energy-driven, and may thus indeed help place J1512$+$4422 on the M--$\sigma_{\ast}$ relation \citep[e.g.,][]{Costa2014b}, as argued in Section \ref{42}. 

Compared to the recent star formation activity, the outflow is capable of expelling gas out more efficiently than the recent star formation activity in the current quenching process. Nevertheless, while the maximum outflow velocity reaches $\sim$1740 \kms, the median outflow velocity is $\sim$352 \kms, which does not guarantee that all of the outflowing gas will leave the galaxy and not fall back to the system in the long run. However, even if the outflow cannot escape the galaxy, it may still heat/disturb the interstellar medium (ISM) and/or the circumgalactic medium (CGM) and contribute to the long-term cumulative AGN feedback \citep[e.g.,][]{HarrisonRamosAlmeida}.
As shown in the right panel of Fig. \ref{fig:energetics}, the total kinetic energy outflow rate of J1512$+$4422 falls within the typical range of galaxy-scale quasar-driven outflows observed at lower and similar redshifts. It is comparable to the minimum requirement (0.1--0.5\% of quasar luminosity) for negative quasar feedback that could suppress the star formation as suggested by simulations \citep[e.g.,][]{Choi2012,Richings2018b}.
Note that our current estimates only account for the warm ionized phase of the outflow. Future observations are needed to probe the neutral and molecular phases of the outflow and obtain the total outflow energetics. 
Overall, the quasar-driven outflow in our object is capable of suppressing/quenching the current star formation activity. 

Moreover, compared to J1512$+$4422, J2236$+$0032 exhibits a faster (1767$\pm{78}$ \kms) outflow, lower sSFR ($\lesssim$0.01 Gyr$^{-1}$) within the last 10 Myr and similar quasar luminosity, BH mass and Eddington ratio based on the FS spectrum. While we currently have no IFU data to spatially resolve the outflowing gas and measure the outflow energetics robustly in this object, the feedback from this outflow is likely at least as powerful as the one in J1512$+$4422 and thus responsible for suppressing/quenching the star formation in the same way.

\begin{figure*}[!htb]
    \centering
    \begin{minipage}{0.49\textwidth}
    \includegraphics[width=1\linewidth]{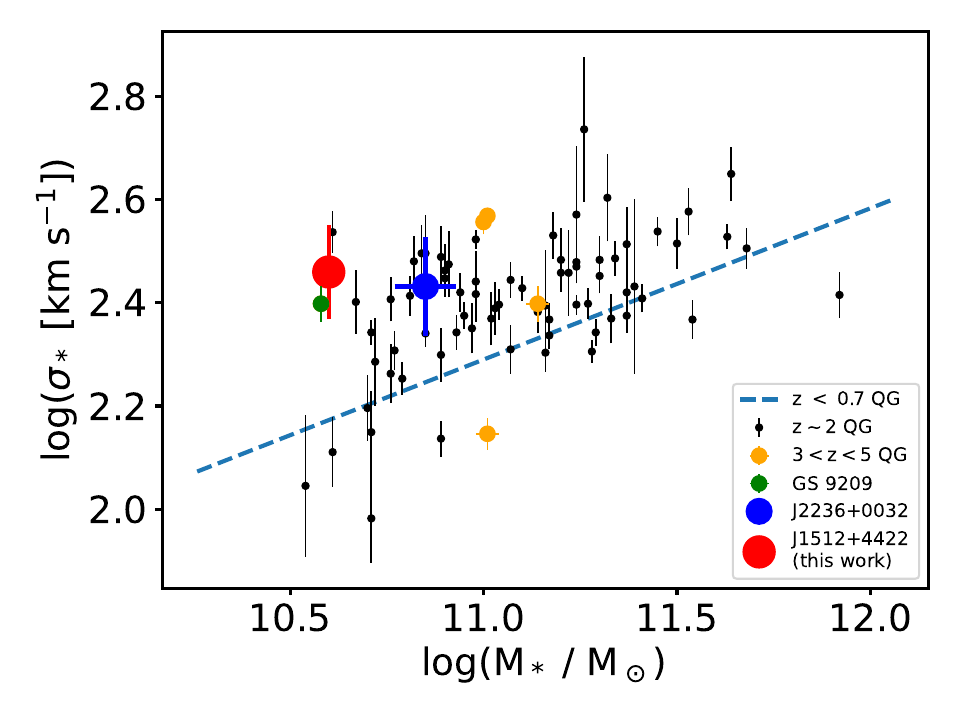}
    \end{minipage}    
    \begin{minipage}{0.48\textwidth}
    \includegraphics[width=1\linewidth]{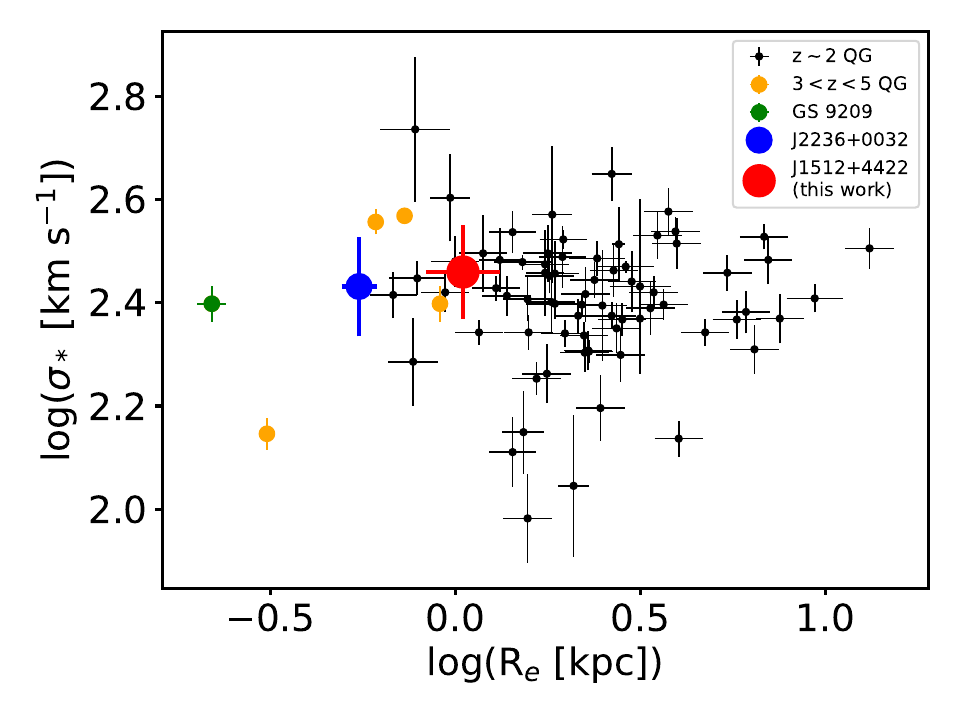}
    \end{minipage}
    \caption{Stellar velocity dispersions as a function of stellar masses (left) and effective radii (right) for J1512$+$4422 (red) and J2236$+$0032 (blue). As a comparison, the 3$<$z$<$5 \citep{Carnall2024} and $z \sim$ 2 \citep{Sande2013} quiescent/post-starburst galaxies are shown in orange and black data points, respectively. The $z \sim$ 4.7 quiescent galaxy GS-9209 \citep{Carnall2023} is shown in green. The best-fit values for quiescent galaxies at $z<$ 0.7 in \citet{Zahid2016} are indicated by the blue dashed line.}
    \label{fig:sigmare} 
\end{figure*}

\section{Connection with $z>2$ Quiescent/Post-starburst Galaxies}
\label{6}

As the host galaxies of both J1512$+$4422 and J2236$+$0032 are quenching, we now explore how they are connected to the $z\gtrsim2$ quiescent/post-starburst galaxies in general. For the effective radius ($R_{\rm e}$) of the two objects, we quote the measurements based on the JWST/NIRCam images from \citet{Onoue2025}.
As shown in Fig. \ref{fig:sigmare}, in the planes of \sig\ versus \mstar\ and $R_{\rm e}$, J1512$+$4422 falls within the boundaries of the regions occupied by $z\sim$ 2--5 quiescent/post-starburst galaxies \citep{Sande2013, Carnall2024}. 
At its \mstar, the \sig\ of J1512$+$4422 is comparable to the highest value observed in $z\sim$ 2--5 quiescent/post-starburst galaxies and higher than the typical value of $z<$ 0.7 quiescent galaxies studied by \citep{Zahid2016}. While on the lower end of both the $R_{\rm e}$ and \mstar\ ranges of all $z\sim$ 2--5 quiescent/post-starburst galaxies examined, J1512$+$4422 exhibits a \sig\ comparable to those with \mstar\ and $R_{\rm e}$ even an order of magnitude higher. Despite of its larger size, J1512$+$4422 has almost the same \mstar\ and \sig\ as GS-9209 at $z\sim4.7$, one of the earliest high-$z$ quiescent galaxies observed so far with a SFR averaged over the past 100 Myr consistent with zero \citep{Carnall2023}.
J2236$+$0032 is located in regions similar to J1512$+$4422 with higher stellar mass and smaller effective radius. Furthermore, GS-9209 also hosts a broad line AGN. As shown in the left panel of Fig. \ref{fig:sigmare}, it is also located on the local $M_{\rm BH}$--\sig\ relation and very close to both J1512$+$4422 and J2236$+$0032. This again suggests a close link between the two quasars and high-$z$ quiescent galaxies. 

Overall, the host galaxies of both J1512$+$4422 and J2236$+$0032 already resemble the $z\sim$ 3--5 quiescent/post-starburst galaxies and the more compact, lower mass population among all $z\sim2$ ones. The two could in principle evolve into the earliest massive quiescent galaxies like GS-9209, once their quasars dim (i.e, decreasing \edd) and star formation is fully quenched. The low \edd\ of J1512$+$4422 ($\sim$0.016) and J2236$+$0032 ($\sim$0.12), and that they are located on the local M--\sig\ relation are both consistent with this picture. Such connection between early quasars and quiescent/post-starburst galaxies has also been suggested by previous studies \citep[e.g.,][]{Onoue2025,Ding2025ApJ,Liu2026a}.

\section{Conclusion}
\label{7}

In this paper, we report new JWST NIRSpec/IFU observations of the quasar J1512$+$4422 at $z\sim6.18$ from the \textit{Aether} survey. 
Our main results are summarized below:

\begin{itemize}

\item 

Based on our new data, J1512$+$4422 has an \logLbol\ of $\sim$45.1 from the 5100 \AA\ continuum luminosity, and an \mbh\ of $\sim$8.9$\times$10$^{8}$ \msun\ ($\sim$6.2$\times$10$^{8}$ \msun) and an \edd\ of $\sim$0.011 (0.016) from the broad \ha\ (\hb) emission line. These are broadly consistent with previously reported values. We measure a stellar velocity dispersion $\sigma_\ast$ $\simeq$ 288$\pm{60}$ \kms\ based on the Balmer absorption lines. This value is larger than the previously reported value ($<$190 \kms), which is based on a deeper FS spectrum with lower spectral resolution and different constraints on the stellar continuum contribution in the spectrum. Meanwhile, the emission line width of the kinematically quiescent gas (i.e., narrow and near the systemic velocity) measured from our data ($\sim$86$\pm{9}$ \kms) is much smaller than $\sigma_\ast$, suggesting that the gas velocity dispersion is not a good surrogate for the stellar one in this object. Our new results indicate that J1512$+$4422 is already on the local M--$\sigma_\ast$ relation just 900 Myr after the Big Bang, similar to the other quasar, J2236$+$0032, with stellar absorption lines detected in \citet{Onoue2025}.

\item 
In J1512$+$4422, we detect a quasar-driven outflow with a velocity of 478$\pm{14}$ \kms\ in the nuclear region. The outflow likely extends to $\sim$3.2 kpc in projection and has a median velocity of 352$\pm{75}$ \kms, whereas the kinematically quiescent gas within the system has velocities with absolute values $\lesssim$50 \kms. 
The outflow is launched about $\sim$ 9 Myr ago, consistent with the time time scale ($\sim$10 Myr ago) of the recent quenching process (sSFR $\lesssim$0.2 Gyr$^{-1}$).
The total mass outflow rate (97.7$^{+97.7}_{-78.2}$ \msunyr
) is larger than the SFR of the system within the last 10 Myr when assuming either a delay-$\tau$ SFH (0.9$^{+3.8}_{-0.8}$ \msunyr) or a non-parametric SFH (4.3$^{+5.8}_{-3.7}$ \msunyr).
The total momentum outflow rate is $6.7^{+6.7}_{-5.4}\times$ the momentum flux from the quasar radiation.
The total kinetic energy outflow rate is $0.6^{+0.6}_{-0.5}$\% of the quasar bolometric luminosity, reaching the lowest threshold for negative quasar feedback as suggested by theoretical models. Overall, this quasar-driven outflow is capable of suppressing or quenching the star formation activity within this system. It could also help place J1512$+$4422 on the local M--$\sigma_\ast$ relation. Such a scenario is also plausible for J2236$+$0032 with an apparently faster outflow.

\item 

In the \mstar--\sig\ and $R_e$--\sig\ planes, both J1512$+$4422 and J2236$+$0032 fall within the regions occupied by $z\sim$ 2--5 quiescent/post-starburst galaxies and are located close to the $z\sim 4.7$ quiescent galaxy, GS--9209 and other $z\gtrsim3$ ones. J1512$+$4422, J2236$+$0032 and GS--9209 fall in the same region of the local $M_{\rm BH}$--$\sigma_\ast$ relation. These suggest that both quasars could evolve into quiescent galaxies like GS--9209 and other $z\gtrsim3$ ones once their quasar fade away and star formation quenched.

Our results point to a tantalizing picture for the evolution of, at least a certain population of, quasar host galaxies within the first billion years after the Big Bang: 
The growth of the SMBHs and their host galaxies are tightly connected via effective feedback, with galaxy-scale outflows representing one critical channel. Such feedback both helps place these objects on the local M-$\sigma_{\ast}$ relation and suppresses/quenches the star formation within their host galaxies very effectively. These quasar host galaxies are among the most promising progenitors of high-$z$ massive quiescent galaxies found by recent studies.

\end{itemize}

\newpage

\appendix

\begin{figure*}[!htb]
    \centering
    \begin{minipage}{0.31\textwidth}
    \includegraphics[width=1\linewidth]{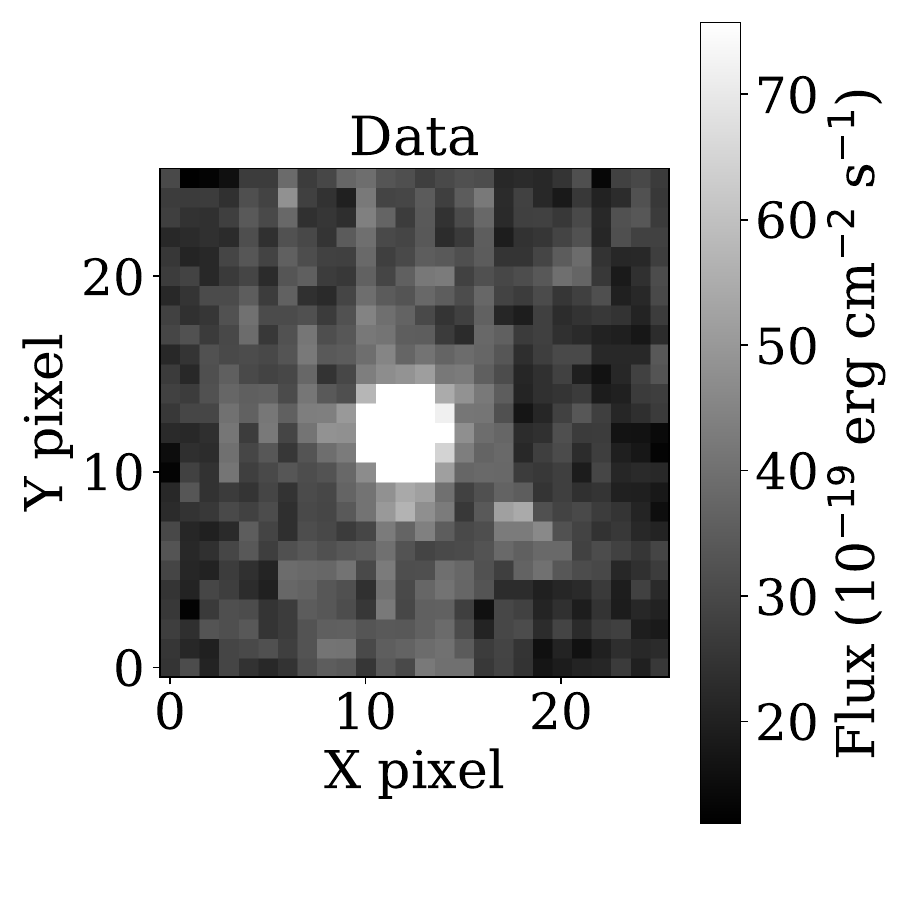}
        \end{minipage}
    \begin{minipage}{0.31\textwidth}
    \includegraphics[width=1\linewidth]{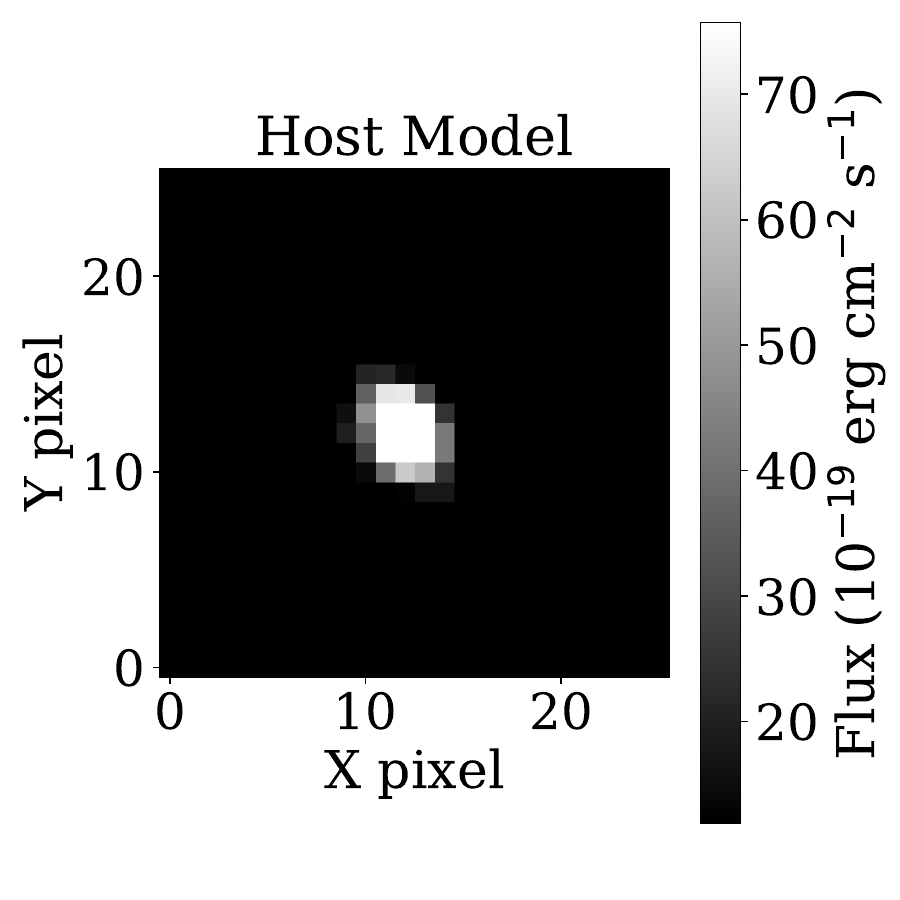}
        \end{minipage}
    \begin{minipage}{0.31\textwidth}
    \includegraphics[width=1\linewidth]{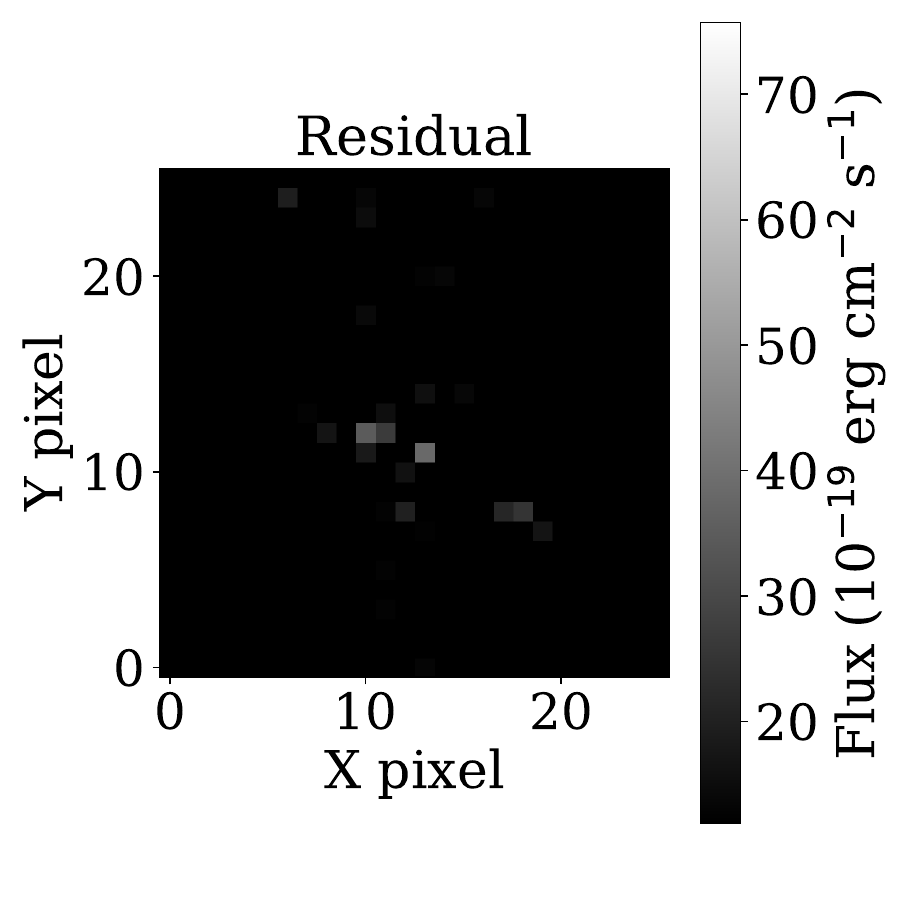}
        \end{minipage}

    \begin{minipage}{0.31\textwidth}
    \includegraphics[width=1\linewidth]{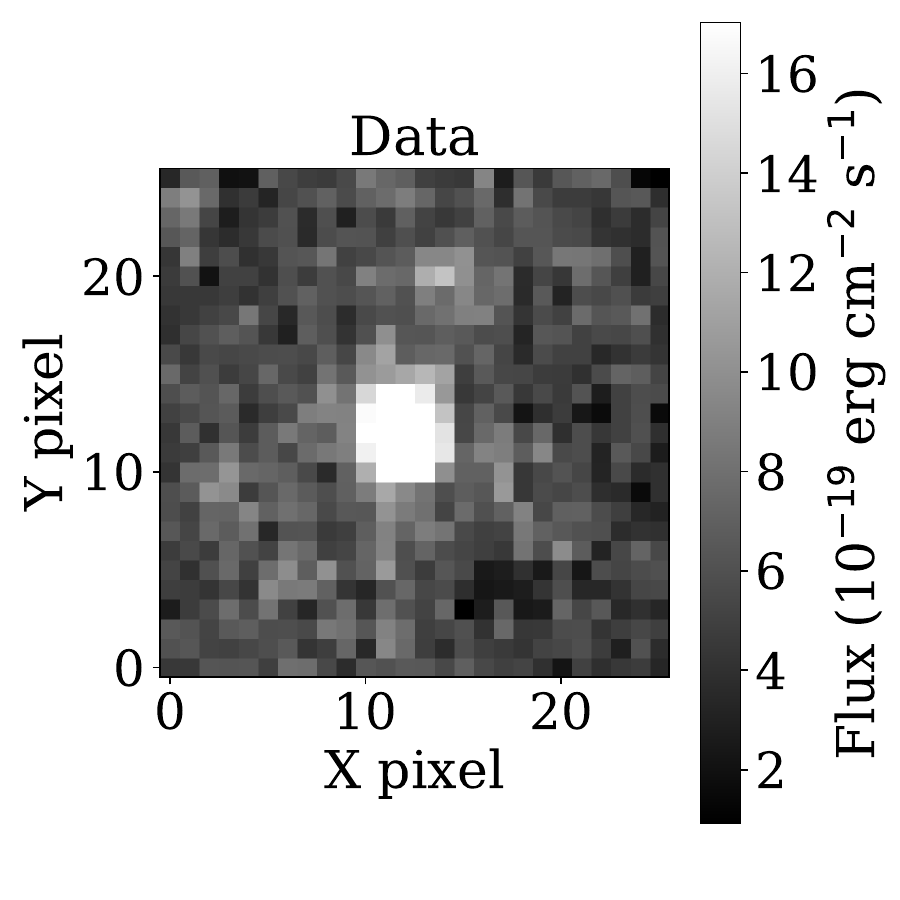}
        \end{minipage}
    \begin{minipage}{0.31\textwidth}
    \includegraphics[width=1\linewidth]{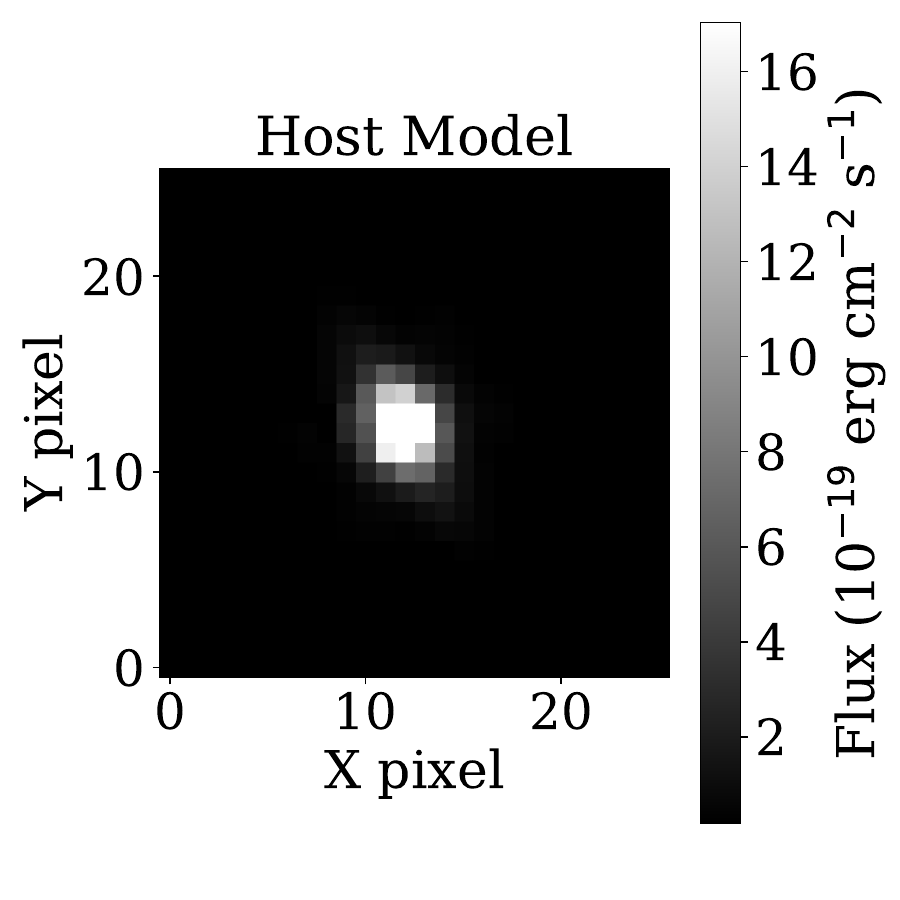}
        \end{minipage}
    \begin{minipage}{0.31\textwidth}
    \includegraphics[width=1\linewidth]{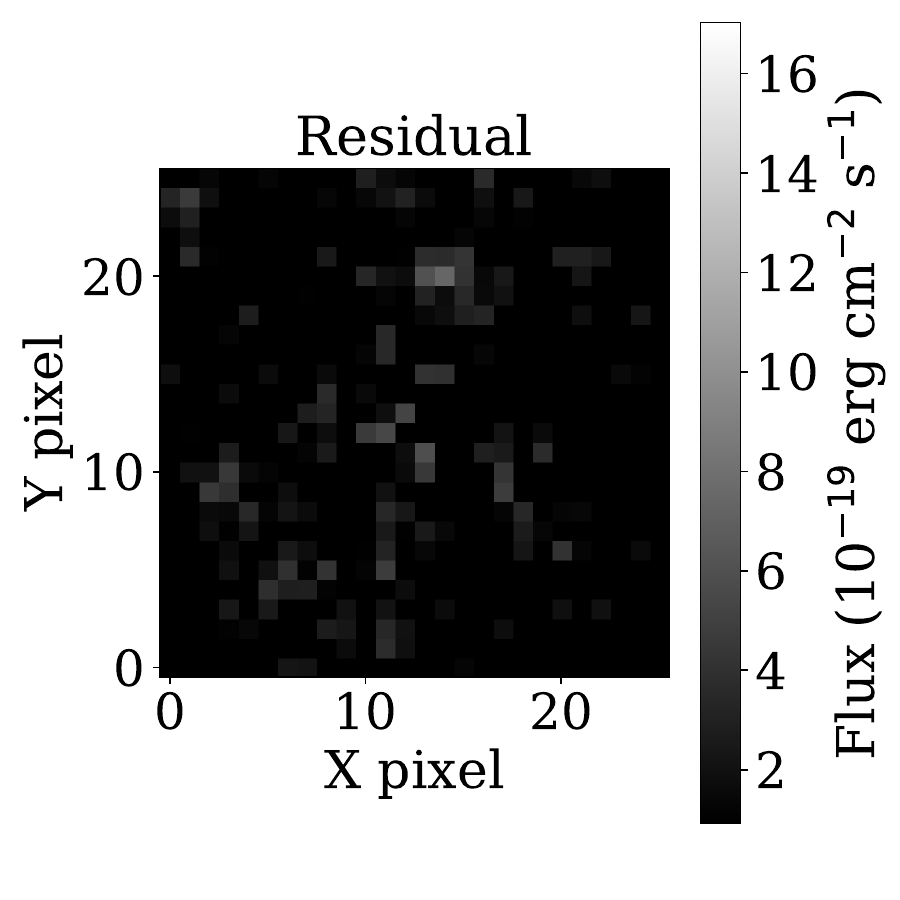}
        \end{minipage}
    \caption{Imaging decomposition results with GALFIT for the 4050--4450 \AA (top) and 5050--5150 \AA (bottom) images from our IFU data. In the decomposition, we use a point source component for the quasar, a S\'ersic component for the host galaxy, and a constant sky background. In each row, the original image, best-fit host galaxy model and residual are shown rom left to right, respectively. In each panel, North is up and East is to the left.}
    \label{fig:galfit}
\end{figure*}

\section{Image Decomposition of the IFU data of J1512$+$4422 with GALFIT.}
\label{galfit}

We use GALFIT \citep{galfit} to decompose the continuum images integrated over 4050--4450 \AA\ and 5050--5150 \AA\ from the IFU data and obtain the corresponding quasar continuum flux free of emission lines. In the decomposition, we use a point source component for the quasar, a S\'ersic component for the host galaxy, and a constant background for the sky background unaccounted for, as shown in Fig. \ref{fig:galfit} in the Appendix. The point spread function (PSF) model is built from the NIRSpec/IFU data of the standard star TYC 4433-1800-1 (proposal ID: 1128) over the same wavelength ranges. The star observations were reduced with the same procedures as our object. The S\'ersic index is fixed to 3, following \citet{Onoue2025}. This choice is motivated by a morphology study of high-redshift quiescent galaxies from \citep{Ito2024}. Nevertheless, changing S\'ersic index to other values (e.g., 1, 2 and 4) does not affect the goodness-of-fit and final results significantly.

The continuum images, best-fit host galaxy models and final residuals from the image decompositions are shown in Fig. \ref{fig:galfit}.
The host galaxy model has effective radii R$_e$ of 0\farcs{071}$\pm{0\farcs{010}}$ (0.42$\pm{0.05}$ kpc) at 4050--4450 \AA\ and 0\farcs{075}$\pm{0\farcs{009}}$ (0.40$\pm{0.06}$ kpc) at 5050--5150 \AA. The best-fit results suggest that the quasar contributes $\sim$50.9\% at 4050--4450 \AA{} and $\sim$61.1\% at 5050--5150 \AA\ within the $r=$0\farcs{1} aperture adopted to extract the nuclear spectrum, respectively. We then fit a power-law model to the obtained quasar flux within the two spectral windows and obtain a best-fit power-law index of $\sim$$-$1.29, which is within the typical range observed in quasars \citep{Selsing2016} and thus confirms that the fit is reliable. This best-fit power-law model is used as the quasar power-law continuum component in the fitting of the nuclear spectrum described in Section \ref{sec:31}.
In the analyses above, we adopt the IFU data cube resampled to 0\farcs{1} spatial pixel (spaxel) size to increase the S/N of continuum in individual spaxels. However, using the original data cube with 0\farcs{05} spaxel size gives consistent results despite of larger uncertainties.

Note that the best-fit host galaxy model based on the JWST/NIRCam F356W imaging from \citet{Onoue2025} is more extended (with effective radius R$_e=$ 0\farcs{19}$\pm${0\farcs{04} or 1.05$\pm{0.23}$ kpc) than our best-fit values. This is likely due to two reasons: i) the F356W image includes both continuum and strong emission lines (i.e., \oiiiab, \hb) whereas the IFU images only include the continuum, so the more extended part of F356W image is dominated by the emission lines. Consistently, the size of extended line emission measured from the IFU data is indeed comparable with the size of the host galaxy based on F356W image; ii) our IFU observation (exposure time $\sim$45 min) is shallower than the imaging (exposure time $\sim$54 min) and fails to capture the more extended faint continuum emission. 

\begin{figure*}[!htb]
    \centering
    \begin{minipage}{0.49\textwidth}
    \includegraphics[width=1\linewidth]{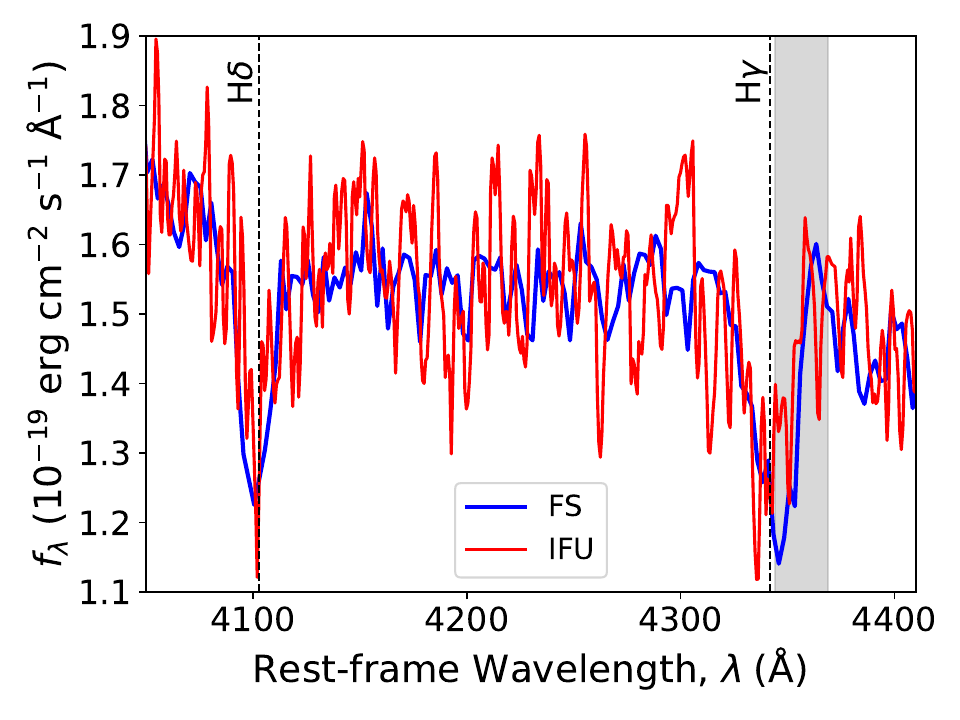}
    \end{minipage}    
    \begin{minipage}{0.48\textwidth}
    \includegraphics[width=1\linewidth]{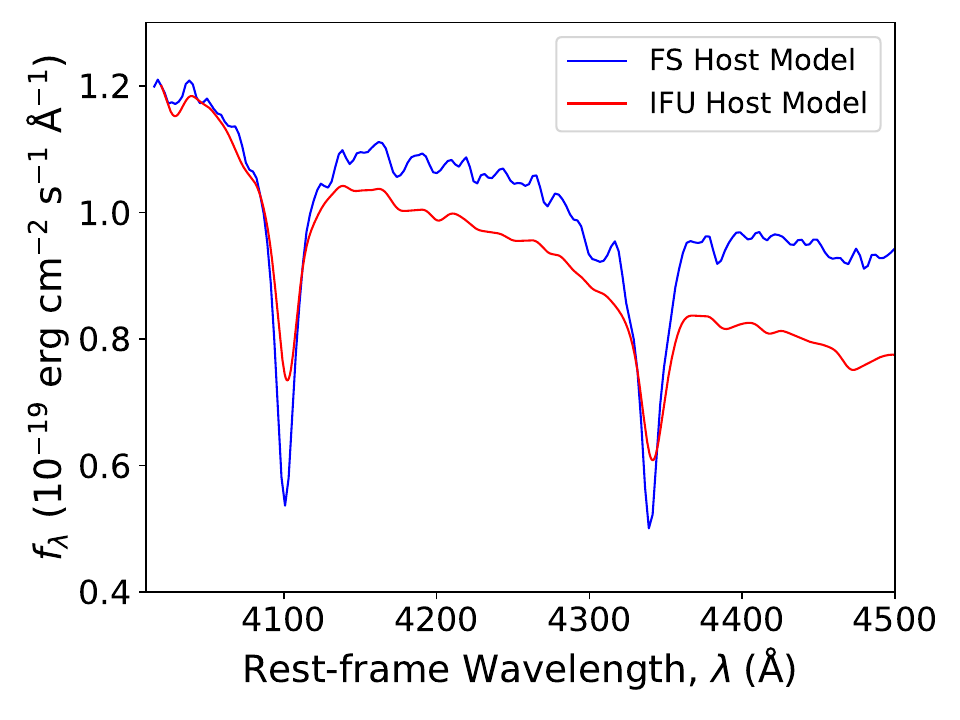}
    \end{minipage}
\caption{\textbf{Left:} Comparison of the IFU (red) and FS (blue) spectra near the stellar absorption features. The gray vertical band indicates the spectral range excluded from the fit to determine \sig\ in \citep{Onoue2025}. The IFU spectrum is rescaled to the same median flux density over the 4400--4600 \AA\ range of the FS spectrum. \textbf{Right:} Comparison of the best-fit host galaxy model of the IFU (red) and FS (blue) spectra. The former is rescaled to the same median flux density over 4010--4020 \AA. In all panels, North is up and East is to the left.} 
    \label{fig:comparespec} 
\end{figure*}

\section{Comparisons of IFU and FS Spectra of J1512$+$4422}
\label{append2}

Comparisons of the IFU and FS spectra and the corresponding best-fit host galaxy models are shown in Fig. \ref{fig:comparespec}.

\begin{acknowledgments}

WL thanks John Silverman for his constructive suggestions.
The IFU data are available on the Mikulski Archive for Space Telescopes (MAST) at the Space Telescope Science Institute, which can be accessed via \dataset[10.17909/mnjk-m858]{https://doi.org/10.17909/mnjk-m858}. WL acknowledges support from NASA through STScI grants JWST-Survey-3428 and HST-GO-17758. RD acknowledges support from the PRORIS 2025 ``CosmoWebb'' and the INAF minigrant 2024: ``The interstellar medium at high redshift''.
MO is supported by the Japan Society for the Promotion of Science (JSPS) KAKENHI Grant Number 24K22894 and 26K07155.
CM acknowledges support from Fondecyt Iniciacion grant 11240336 and the ANID BASAL project FB210003. 

\end{acknowledgments}

\begin{contribution}
WL conceived the project, carried out the data reduction and analysis, and wrote the manuscript. EF is the PI of the \textit{Aether} survey program. EF, XF, RD and MO provided constructive suggestions for the data analysis and interpretation starting from the early stage. MO provided data and measurements from \citet{Onoue2025}. HZ provided the results from the \textit{Trinity} model. All authors reviewed and provided constructive comments on the manuscript.
\end{contribution}

\facilities{JWST}

\software{astropy \citep{Ast2013, Ast2018}, reproject (\url{https://doi.org/10.5281/zenodo.7584411}), \qtdfit\ \citep{q3dfit}, pPXF \citep{ppxf}, BAGPIPES \citep{BAGPIPES}.}

\bibliography{J1512}{}
\bibliographystyle{aasjournalv7}



\end{document}